\documentclass[a4paper,11pt]{article}
\pdfoutput=1 

\usepackage{jheppub} 
\usepackage{lmodern}
\usepackage{tikz}
\usepackage{xcolor}
\usetikzlibrary{decorations.pathmorphing}
\tikzset{snake it/.style={decorate, decoration=snake}}
\usepackage{slashed}
\usepackage[]{mdframed}
\usepackage[T1]{fontenc} 
\newcommand{\be}{\begin{equation}}
\newcommand{\bea}{\begin{eqnarray}}
\newcommand{\ee}{\end{equation}}
\newcommand{\eea}{\end{eqnarray}}

\title{\boldmath 
Holographic Krylov Complexity for Charged, Composite and Extended Probes
}
\author[a]{Horatiu Nastase,}
\author[b]{ Carlos Nunez}
\author[c]{and Dibakar Roychowdhury}
\affiliation[a]{Instituto de F\'\i sica Te\'orica, UNESP-Universidade Estadual Paulista R. Dr. Bento T. Ferraz 271, Bl. II, Sao Paulo 01140-070, SP, Brazil}
\affiliation[b]{Centre for Quantum Fields and Gravity, Department of Physics, Swansea University, Swansea SA2 8PP, United Kingdom}
\affiliation[c]{Department of Physics, Indian Institute of Technology Roorkee,\\Roorkee 247667, Uttarakhand, India}
\abstract{We study a proposed holographic representation of spread/Krylov complexity for states or
operators dual to charged, composite and extended probes. Our analysis builds on the relation
between the rate of spread complexity and the proper momentum of an infalling bulk excitation.
A conserved Noether charge requires particular care: unitary spread complexity satisfies
$\dot{\mathcal C}_{K}(0)=0$, and therefore the velocity conjugate to a fixed charge cannot be
counted as an independent spreading direction. We consequently eliminate cyclic coordinates by
a partial Legendre transform and compute the proper momentum from the fixed-charge Routhian.

We first apply this prescription to an $R$-charged particle in AdS$_5\times S^5$. The charge
enters through the reduced effective mass and energy, while the resulting complexity has the
required quadratic onset. We next study probes that are pointlike from the boundary perspective
but possess non-trivial bulk structure. These are baryon-vertex configurations and giant gravitons. For the
giant graviton the fixed-charge prescription retains the effects of the Born--Infeld and
Wess--Zumino couplings without reintroducing the cyclic angular velocity into the proper
momentum. We also analyse a fundamental string falling in AdS while stretched along a boundary
spatial direction as a collective model of a non-local excitation.

Within the rigid-probe Poincar\'e-AdS examples considered here, the proper-momentum rate is
linear at late times, whereas charges, compositeness and spatial extension affect its
normalisation and its subleading or intermediate-time structure. We interpret this as a robust
pattern in the tested class (not as a general universality theorem). The fixed-charge
Routhian provides a candidate bulk counterpart of sector selection. An explicit
symmetry-resolved boundary Lanczos construction remains an open problem.}
\begin{document} 
\maketitle
\flushbottom
\section{Introduction and General Idea}
%
%
%
%
%
%
%
%
%
Quantum complexity has become a useful meeting point between quantum information,
many-body dynamics, quantum chaos and holography. In recent years it has become
increasingly clear that the word ``complexity'' covers several related but genuinely
different notions, ranging from circuit and Nielsen complexity to tensor-network,
path-integral and holographic proposals. Among these, Krylov complexity has emerged
as an especially natural diagnostic of operator growth. Starting from a reference operator
and repeatedly acting with the Liouvillian, one generates a distinguished orthonormal basis
through the Lanczos algorithm; the corresponding Krylov complexity measures how far the
time-evolved operator has spread along this dynamically generated chain. One of its main
appeals is precisely that the basis is not imposed externally, but is instead fixed by the
Hamiltonian and by the operator under consideration. In this sense, Krylov complexity
provides a canonical notion of spreading, well suited to questions about scrambling,
universality and the detailed operator-dependence of quantum evolution
\cite{Parker:2018yvk, 
Barbon:2019wsy, 
Avdoshkin:2019trj,
Dymarsky:2021bjq,
Caputa:2021sib,
Balasubramanian:2022tpr,
Baiguera:2025dkc,
Rabinovici:2025otw, Nandy:2024evd}.

The modern development of the subject was strongly shaped by the operator-growth
perspective of \cite{Parker:2018yvk}, who proposed that in generic many-body systems the
Lanczos coefficients display an asymptotically linear growth. This viewpoint connected the
structure of the Krylov chain to universal properties of chaotic dynamics and stimulated a
rapidly expanding literature on late-time growth, bounds, integrability, quantum field
theory and geometric formulations of Krylov space
\cite{Parker:2018yvk,Barbon:2019wsy,Avdoshkin:2019trj,Dymarsky:2021bjq,
Caputa:2021sib,Balasubramanian:2022tpr,Rabinovici:2020ryf,Rabinovici:2022beu, Baggioli:2024wbz}.
A particularly important lesson from this body of work is that Krylov complexity is not
merely a coarse measure of chaos. It is a probe of how a specific operator explores the
available Hilbert-space directions, and it is therefore sensitive to the initial operator, to
symmetries, to conserved quantum numbers, and to the detailed dynamical sector in which
the evolution takes place. This sensitivity is especially attractive from a holographic point of
view, where one would like to understand which bulk observables capture not only the
overall growth of complexity, but also its dependence on the nature of the probe.

On the holographic side, the earliest and most influential proposals for complexity were the
complexity-volume and complexity-action conjectures, together with their many
generalisations and refinements. In these frameworks, one relates the complexity of the
boundary state to an extremal volume or to the action evaluated on a Wheeler--DeWitt
patch, thereby extending to complexity the geometric spirit that proved so fruitful in other
holographic observables \cite{Susskind:2014rva,Brown:2015bva,
Baiguera:2025dkc}. More recently, a complementary line of development
has aimed at connecting spread or Krylov complexity to bulk kinematics more directly \cite{Balasubramanian:2022tpr, Balasubramanian:2025xkj}. In
particular, Krylov spread complexity admits an explicit bulk realisation in JT gravity in suitable
limits, while for local operator excitations in AdS$_3$/CFT$_2$ the proposal of
\cite{Caputa:2024sux} relates the rate of spread complexity to the proper radial momentum
of an infalling probe \cite{Rabinovici:2023yex,Caputa:2024sux}. This sharpened the idea
that bulk motion can encode the boundary growth of complexity in a precise and calculable
manner.

There is a necessary consistency condition on any proposed bulk representation of spread
complexity. Given a normalised seed $|K_0\rangle$, the Lanczos expansion
\begin{equation}
e^{-iHt}|K_0\rangle=\sum_{n\geq 0}\varphi_n(t)|K_n\rangle,
\qquad
\mathcal C_K(t)=\sum_{n\geq0}n|\varphi_n(t)|^2,
\label{eq:Krylov-definition-revised}
\end{equation}
implies $\varphi_1(t)=-ib_1t+\mathcal O(t^2)$ and $\varphi_n(t)=\mathcal O(t^n)$.
For unitary evolution generated by a Hermitian Hamiltonian, the corresponding Jacobi matrix
may be chosen real, so that $|\varphi_n(t)|^2$ and $\mathcal C_K(t)$ are even functions of
time. Hence
\begin{equation}
\mathcal C_K(t)=b_1^2t^2+\mathcal O(t^4),
\qquad
\dot{\mathcal C}_K(t)=2b_1^2t+\mathcal O(t^3),
\qquad
\dot{\mathcal C}_K(0)=0.
\label{eq:universal-short-time}
\end{equation}
This is a structural property of the Krylov construction, rather than a model-dependent
expectation \cite{Huh:2023xtr,Fan:2022xaa,Muck:2026top}. A bulk proper momentum containing
a non-zero constant at the reference time therefore cannot directly represent the standard
unitary Krylov-complexity rate. In the examples below this issue arises when the probe moves
along a cyclic direction carrying a conserved Noether charge. Our prescription is to fix the
charge first by a partial Legendre transform and to compute the proper momentum from the
resulting Routhian \cite{Chatzis:2026ekd}. Throughout the paper we use the orientation-independent
convention (this makes the complexity of falling probes grow with time)
\begin{equation}
\dot{\mathcal C}(t)=\kappa |P_y(t)|,\qquad \kappa>0,\qquad t>0,
\label{eq:absolute-momentum-prescription}
\end{equation}
where $\kappa$ allows for a convention-dependent normalisation.

The purpose of this paper is to explore how far this proposal can be pushed when the probe
is no longer a completely structureless particle. Our viewpoint is that this question is
particularly natural in holography. A local operator may be represented in the bulk by a
pointlike object, but that object can still carry non-trivial internal data, like conserved charges,
worldvolume fluxes, induced couplings, or a composite origin in terms of higher-dimensional
branes. Conversely, one may also consider genuinely extended probes, whose dual boundary
interpretation is manifestly non-local. It is then natural to ask which features of the
Krylov-complexity story are universal, which are sensitive to the operator being probed, and
which are artifacts of the simplest point-particle approximation.

\subsection{General idea of this paper}

Most holographic discussions of complexity growth based on infalling probes use objects
whose dynamics is effectively pointlike from the boundary point of view. In the simplest
setups one studies a massive particle falling along a radial geodesic, and the proper
momentum of this probe reproduces the expected complexity growth \cite{Caputa:2024sux, Fan:2024iop, He:2024pox, Jeong:2026iac}. This is both elegant
and physically transparent. But it also leaves open an important issue: whether the same
dictionary continues to hold when the operator is still local for the field-theory observer,
yet carries an internal structure that is non-trivial in the bulk.

The first aim of this work is to study precisely this class of probes. We consider, on the one
hand, a particle carrying a conserved $R$-charge, whose motion involves both the AdS radial
direction and the internal space, in the spirit of BMN-like sectors
\cite{Berenstein:2002jq}. We then turn to probes that are pointlike for the boundary
observer but arise from higher-dimensional branes with non-trivial worldvolume Physics, like
baryon-vertex configurations \cite{Witten:1998xy} and giant gravitons
\cite{McGreevy:2000cw,Grisaru:2000zn}. These are natural laboratories in which to test
the dependence of spread complexity on conserved quantum numbers, compositeness,
induced charges and Wess--Zumino couplings, while remaining in a kinematical regime that
is still close to the original point-particle proposal.

The examples are selected so as to vary the nature of the probe while keeping the central
Poincar\'e-AdS geometry fixed. The ordinary particle supplies the baseline; the $R$-charged
particle isolates a conserved Noether sector. The baryon vertex adds compositeness, attached
strings and induced charge. The giant graviton introduces wrapped-brane dynamics and a
Wess--Zumino coupling. Finally, the spatially stretched fundamental string changes the effective
radial action of a non-local excitation. This organisation allows us to distinguish effects
of probe structure from those of the background geometry. It also makes clear why the
late-time agreement found below is evidence within a controlled class, rather than a proof
of universality.

Our second aim is to move beyond this effectively pointlike sector and consider a genuinely
extended probe, namely a fundamental string falling in AdS while stretched along a spatial
direction. From the boundary viewpoint, this is intended to model a non-local operator.
This case is conceptually more delicate, and for that reason also more interesting. Following
the prescription advocated in \cite{Caputa:2024sux}, we assume that a suitably defined
generalised proper momentum still captures the rate of spread complexity. For rigid or
consistently truncated configurations, this is a natural semiclassical working hypothesis.
Nevertheless, its status should be stated with some care when extended objects are involved.

Indeed, several possible criticisms become relevant in the extended case. First, a string or
brane supports infinitely many internal modes, and a collective-coordinate description need
not capture the full operator growth of the dual non-local excitation. Second, worldsheet or
worldvolume gauge choices, Wess--Zumino terms, endpoint contributions and conserved
charges may affect the definition of the canonical momenta entering the complexity
prescription. Third, for a non-local operator it is less obvious that the resulting quantity
should be identified with the same notion of Krylov complexity that is naturally associated
with local operators. For this reason, throughout the paper we interpret the extended-probe
calculation conservatively: it should be viewed as a controlled proposal for a collective notion
of spread complexity, rather than as a complete first-principles derivation. In our view this is
not a weakness but a useful diagnostic. By comparing structured pointlike probes with
genuinely extended ones, one can isolate which aspects of the momentum/complexity
correspondence are robust and which call for a more refined boundary understanding.

With this perspective in mind, the goal of the paper is not to claim a final holographic
dictionary for every version of Krylov complexity, but rather to enlarge the class of probes
for which the proposal can be tested explicitly. Our emphasis will therefore be on solvable
semiclassical dynamics, on the role of conserved quantities and induced interactions, and on
the contrast between local/composite probes and non-local/extended probes. This also helps
to identify the main open problem on the field-theory side. That is, to isolate the boundary sectors
and operators for which an independent Krylov or spread-complexity computation can be
matched to the bulk results discussed here.

The paper is organised as follows. In section \ref{CONSRSR} we study a particle falling in
AdS$_5$ while carrying a conserved $R$-charge. We separate the unreduced momentum of the
full bulk trajectory from the proper momentum of the fixed-charge Routhian, and identify
only the latter with the candidate Krylov-complexity rate. In section \ref{sec:BV} we turn
to baryon-vertex-like operators, which are composite but effectively pointlike probes, and
derive their dynamics and associated proper momentum, including the effect of the attached
fundamental strings. In section \ref{complexitygiant} we consider a falling giant graviton.
Its Born--Infeld and Wess--Zumino couplings provide a second, more stringent application of
the fixed-charge prescription. Section \ref{complexityextended-string} studies a genuinely
extended fundamental string as a collective model of the spread of a non-local excitation.
We conclude in section \ref{section-concl}. Appendix \ref{appendixb} extends the
baryon-vertex analysis to AdS$_3\times S^3\times CY_2$, while appendix \ref{appendixa}
rederives the baryon-vertex dynamics using a Routhian formulation. Appendix \ref{appendixc}
gives the fixed-charge short-time expansion for the giant graviton, and appendix
\ref{appendix-fixedcharge-routhian} collects additional details of the Routhian prescription.
\section{Conserved $R$-charge and the fixed-charge prescription}\label{CONSRSR}

In this section we consider a massive particle that falls radially in AdS$_5$ and carries
angular momentum $J$ along an equator of $S^5$. The angular coordinate is cyclic, so $J$
labels a conserved $R$-charge sector. Our aim is to determine the proper momentum appropriate
to that {\it fixed charge sector}. We first display the full bulk trajectory, but apply the complexity
prescription only after eliminating the cyclic angle by a partial Legendre transform. This
order of operations is required by the universal short-time law
$\mathcal C_K(t)=b_1^2t^2+\mathcal O(t^4)$ in eq.~\eqref{eq:universal-short-time}.
The configuration  on the QFT side may be viewed as a BMN-like excitation built from adjoint fields and
covariant derivatives \cite{Berenstein:2002jq}. The generalised proper-momentum constructions
of \cite{Fatemiabhari:2025poq,Fatemiabhari:2026goj,Fatemiabhari:2026rob,
Roychowdhury:2026eds} will be useful below, but a conserved cyclic charge requires the
additional fixed-sector reduction described in \cite{Chatzis:2026ekd}.

The ten dimensional metric and five form are
\begin{eqnarray}
& & ds^2=\frac{r^2}{l^2}\left(-dt^2+d\vec{x}_3^2 \right) +\frac{l^2 dr^2}{r^2}+ l^2d\Omega_5^2,\label{AdS5metrica}\\
& &F_5=N_c(1+*) \text{Vol}_{S^5}.\nonumber
\end{eqnarray}
The five sphere has line element
\begin{align}
d\Omega_5&=d\theta_1^2+\sin^2\theta_1 d\theta_2^2+ \sin^2\theta_1 \sin^2\theta_2 d\theta_3^2+  \sin^2\theta_1 \sin^2\theta_2 \sin^2\theta_3d\theta_4^2  \nonumber\\& + \sin^2\theta_1\sin^2\theta_2 \sin^2\theta_3\sin^2\theta_4 d\psi^2.\nonumber
\end{align}
Let us consider a point particle probing AdS$_5\times S^5$ in the configuration
\begin{equation}
r=r(t),~~~\psi=\psi(t).\label{embeddingparticle}
\end{equation}
This is a consistent embedding subjected to the fact that one takes the various angles $\theta_i=\frac{\pi}{2}$, on the five sphere.
The Lagrangian for this particle is
\begin{equation}
L=-m\sqrt{\frac{r^2}{l^2} -\frac{l^2\dot{r}^2}{r^2}-l^2\dot{\psi}^2}.    
\end{equation}
There are two equations of motion and two conserved quantities,
\begin{align}
    &-\frac{d}{dt}\left[\frac{l^2\dot{r}}{r^2 L} \right]= \frac{(r^4+l^4\dot{r}^2)}{l^2 r^3 L},~~~\frac{d}{dt}\left[\frac{\dot{\psi}}{L}\right]=0,\label{eqs-of-motion}\\
    &{-}J=  \frac{m^2 l^2}{L} \dot{\psi}, ~~~{-}H= \frac{m^2 r^2}{l^2 L}.
    \label{conservedcharge}
\end{align}
%
%
One can invert eq.\eqref{conservedcharge} to obtain expressions for $\dot r$ and $\dot\psi$,
\begin{align}
    \dot{\psi}= \frac{J r^2}{Hl^4}, ~~~\dot{r}=\pm\frac{r^2}{Hl^4}\sqrt{H^2 l^4-(J^2+ m^2 l^2)r^2}.
\end{align}
These first order equations solve the second order ones in eq.(\ref{eqs-of-motion}).
The  solution to the first order equations is
\begin{align}
\label{sol}
     r(t)=\pm \frac{H l^2}{\sqrt{H^2 t^2+J^2+l^2 m^2}}~;~ \psi (t)= \frac{J }{\sqrt{J^2+m^2 l^2}}\tan ^{-1}\left(\frac{H t}{\sqrt{J^2+ m^2l^2}}\right).
 \end{align}
Notice that at $t=0$, the particle starts
from the UV-cutoff radial position $r_{UV}=\frac{H l^2}{\sqrt{J^2 +m^2 l^2}}$ and zero initial velocity. Also, for $t=0$ the angle starts from $\psi=0$ with initial angular velocity
$\frac{JH}{J^2+m^2 l^2}$.


We now distinguish two different quantities. The first is the momentum associated with the
full two-dimensional trajectory $(r,\psi)$. It is obtained by introducing
\begin{equation}
dy_{\rm all}^{2}=\frac{l^2}{r^2}dr^2+l^2d\psi^2,
\qquad
\dot y_{\rm all}=\sqrt{\frac{l^2\dot r^2}{r^2}+l^2\dot\psi^2},
\label{properdistance}
\end{equation}
and gives, on the solution \eqref{sol},
\begin{equation}
\mathfrak P_{\rm all}(t)=\frac{1}{l}\sqrt{H^2t^2+J^2}.
\label{eq:unreduced-particle-diagnostic}
\end{equation}
This quantity measures the magnitude of the total semiclassical motion, including the
velocity conjugate to $J$. Since $\mathfrak P_{\rm all}(0)=|J|/l$, it cannot equal the
unitary fixed-sector Krylov-complexity rate. We therefore do not associate it with
$\dot{\mathcal C}$. Its expansions,
\begin{align}
\mathfrak P_{\rm all}(t)&=\frac{|J|}{l}+\frac{H^2}{2|J|l}t^2
+\mathcal O(t^4), && t\to0,\label{cdotatsmallt}\\
\mathfrak P_{\rm all}(t)&=\frac{|H|}{l}|t|+
\frac{J^2}{2|H|l|t|}+\mathcal O(t^{-3}), && |t|\to\infty,
\label{cdotatlarget}
\end{align}
remain useful kinematical diagnostics, but they should not be interpreted as the short-time
or subleading behaviour of a fixed-charge Krylov complexity.
\\
The physical fixed-$J$ prescription is obtained by eliminating $\psi$ before defining the
proper coordinate. Solving $J=\partial L/\partial\dot\psi$ and performing the partial
Legendre transform gives
\begin{equation}
R_J=J\dot\psi-L=\widehat m_J
\sqrt{\frac{r^2}{l^2}-\frac{l^2\dot r^2}{r^2}},
\qquad
\widehat m_J^2=m^2+\frac{J^2}{l^2}.
\label{eq:fixedJ-particle-Routhian-main}
\end{equation}
The charge has become part of the effective mass.  We are now working at {\it a fixed value of $J$}. The reduced proper coordinate and its
momentum are
\begin{equation}
\dot y_J=\frac{l\dot r}{r},
\qquad
P_y^{(J)}=\frac{\partial R_J}{\partial\dot y_J}
=-\frac{\widehat m_J\dot y_J}
{\sqrt{r^2/l^2-\dot y_J^{,2}}}.
\label{eq:fixedJ-particle-Py-main}
\end{equation}
The conserved energy derived from $R_J$ reproduces the radial trajectory in eq.~\eqref{sol}.
Using \cite{Caputa:2024sux} and eq.(\ref{eq:absolute-momentum-prescription}), for the release of the probe from rest at $r_{UV}$ one obtains
\begin{equation}
|P_y^{(J)}(t)|=\frac{|H_J|}{l}|t|,
\qquad
\mathcal C_J(t)-\mathcal C_J(0)
=\frac{\kappa|H_J|}{2l}t^2+\mathcal O(t^4).
\label{cdotJ}
\end{equation}
In particular, $P_y^{(J)}(0)=0$, as required by
eq.~\eqref{eq:universal-short-time}. The charge dependence is retained through
$\widehat m_J$, the initial position and $H_J$; the conjugate angular velocity is not
counted as a direction of spreading within a sector of definite $J$.
\\
For completeness, let us discuss a possible point of contact with symmetry resolved Krylov complexity on the field theory side \cite{Caputa:2025mii,Caputa:2025ozd}.  A genuine symmetry-resolved boundary construction would start from a
charge projector $\Pi_J$ and the normalised seed
\begin{equation}
|\Psi_{0,J}\rangle=
\frac{\Pi_J|\Psi_0\rangle}
{\sqrt{\langle\Psi_0|\Pi_J|\Psi_0\rangle}},
\qquad
H^{\rm QFT}_J=\Pi_JH\Pi_J,
\label{eq:sector-projector}
\end{equation}
and would perform the Lanczos recursion with $H^{\rm QFT}_J$ entirely inside fixed charge Hilbert space \cite{Caputa:2025mii,Caputa:2025ozd}. The present paper
does not construct $\Pi_J$, the sector Lanczos coefficients or the sector Krylov
wavefunctions. The Routhian prescription should therefore be viewed as a candidate bulk
counterpart of working at fixed charge, not as an explicit computation of
symmetry-resolved Krylov complexity. Establishing that correspondence directly in
$\mathcal N=4$ SYM is an important problem for future work.
\\
The coordinates $(t,\vec x,r,y_J)$ have units of length, while the angles are dimensionless.
The parameters $H_J$ and $m$ have units of inverse length, $J$ is dimensionless, and
$P_y^{(J)}$ has units of inverse length. Thus all expressions above are dimensionally
consistent.

Let us now focus on a different 'point particle' probe. The difference is that whilst these probes do not have an extra quantum number (like the R-charge studied above), the operators considered do have an interesting internal structure (being made out of many strongly interacting components).


\section{Complexity for baryon-vertex like operators}\label{sec:BV}
It is known that the Krylov/spread complexity is in principle dependent on the operator whose 'spread' across the Hilbert space is measured, see for example \cite{Craps:2024suj, PG:2025ixk}. In this section, we consider the spread/Krylov complexity  of {\it baryon-vertex like operators} \cite{Witten:1998xy}. In holography, the baryon vertex consists of a D-brane that pulls-back on time and  part of the  internal space. There is also an excited gauge field  on the brane. This differs from the pointlike probes previously considered in the literature. In fact the charge induced on the wrapped D-brane probe is canceled by a set of $N$ F1-strings that end at infinity  (where another D-brane is placed).
In this sense, the baryon-vertex is a  heavy 'composite' operator (the mass is proportional to $N$, the number of F1-strings and the number of colours in the SCFT). From the viewpoint of the field theory observer, it is still a heavy point like object with electric charge and composed by a large number of constituents (non-dynamical quarks). 

Holographically, the complexity is computed by a Dp-brane extended on $[t,\Omega_p]$  that falls along a geodesic in the AdS radial direction, parametrised by $r(t)$. We must consider the action of the $N$-fundamental strings that stretch from the baryon-vertex Dp-brane to the brane at infinity. These F1s contribute to the potential part of the Lagrangian of the system. As a first approximation, we do not add a kinetic term for the F1 strings, and consider them tension-full 'rods'. This approximation is valid as long as  the strings do not fluctuate very much in directions perpendicular to the radial-direction respect to their (radial) stretch. The whole object has a dynamics different from the single particle falling in AdS. It is  interesting to calculate the spread complexity for this  point-like operator.
Before we study the geodesic fall of this composite object, let us write some words about the different descriptions of the baryon vertex, with the purpose of making the forthcoming calculations clearer.
\\
\\
\underline{\bf Comments on baryon vertex}
\\
A baryon is defined as the contraction of $N$ quarks (fundamental fields) with the antisymmetric tensor of SU$(N)$. In the case of a field theory without  fields in the fundamental representation of the gauge group (quarks), the closest object is  a composite of $N$ external quarks. In holography, one works with a  combination of Dp branes and F1 strings. The Dp brane extends along $[t,\Omega_p]$ where $\Omega_p$ belongs to the internal space, making this object look like a particle for the field theory observer. This Dp brane admits a fluctuation described by a gauge field $a_1$ with curvature $f_2$. Aside from the Born-Infeld action of the brane $S_{BI}=T_{Dp} \int d^{p+1}x~ e^{-\Phi}\sqrt{-\det[g+f]}$ we have a Wess-Zumino term (associated with a background Ramond potential $C_{p-1}$)
\begin{equation}
S_{WZ}=-T_{Dp}\int_{(t,\Omega_p)}C_{p-1}\wedge f_2 = T_{Dp}\int_{(t,\Omega_p)} F_p\wedge a_1=N T_{Dp}\int  a_t (t)dt.    
\end{equation}
In the cases we study below $\int_{\Omega_p} F_p= N $. We consider a gauge field on the brane $a_1= a_t(t) dt$, that has zero curvature $f_2=da_1=0$. The action describing this Dp brane is
\begin{equation}
S_{BIWZ}= T_{Dp} \int d^{p+1}x~ e^{-\Phi}\sqrt{-\det[g_{ind}]} + N T_{Dp}\int a_t(t) dt.\label{action-d5}  
\end{equation}
 The equation of motion $\frac{\delta S_{BIWZ}}{\delta a_t}= N=0$, forces us to add $N$-fundamental strings that carry the charge away to satisfy Gauss' law in a compact space $\Omega_p$ (in other words $a_1$ acts as a Lagrange multiplier). The consistent Dp-F1 system has action
\begin{eqnarray}
& &S_{BIWZ-F1}= T_{Dp} \int d^{p+1}x~ 
e^{-\Phi}\sqrt{-\det[g_{ind}]} + N T_{Dp}\int a_t(t) dt +N S_{F1} ,\nonumber\\
&&S_{F_1}=S_{NG}- T_{F1}\int a_t(t) dt\;,\nonumber\\
& &S_{NG}=T_{F1}\int dt  \int_{r(t)}^{r_{UV}}dr\sqrt{-g_{tt }g_{rr}}=T_{F1}\int dt [r_{UV}-r(t)].\label{action-d5-f1}     
\end{eqnarray}

This yields the complete action of the Dp-F1 system
\begin{align}\label{julianalvarez}
    S_{BIWZ-F1}=T_{Dp} \int d^{p+1}x~ e^{-\Phi}\sqrt{-\det[g_{ind}]} +N T_{F1}\int dt [r_{UV}-r(t)].
\end{align}

In our calculation, we allow the probe Dp brane to fall following a radial geodesic in an AdS geometry. We also approximate $S_{F1}\approx\int  dt (r_{UV}-r(t))$ (the F1's extended along the radial direction do not fluctuate as a first approximation) contributing to the potential energy of the system. An alternative view of the baryon vertex (that we do not use here) is in terms of a soliton--a BIon--of the BIWZ action, with a  dynamical gauge field. In this configuration the  charge of F1 is induced on the world volume of the Dp brane. To compare the relevant bibliography explaining the different viewpoints see \cite{Witten:1998xy, Callan:1998iq, Gomis:1999xs, Janssen:2006sc}.

We consider the case of baryon-vertex operators measuring complexity in AdS$_5\times S^5$. In Appendix \ref{appendixb} we extend the treatment to the case of AdS$_3\times S^3\times CY_2$.
In Appendix \ref{appendixa} (for the case of AdS$_5\times S^5$), we present a treatment that reproduces our results above using the Routhian (an intermediate functional between the Lagrangian and the Hamiltonian). This treatment does {\it not} need the gauge field on the brane $a_1$ to have zero curvature $f_2=da_1$. The end result is that of  eq.(\ref{julianalvarez}). 

\subsection{The baryon vertex falling in AdS$_5\times S^5$}\label{sec:BVADS5}
We start studying the baryon vertex in ${\cal N}=4$ SYM. The dual background is AdS$_5\times S^5$. We consider a probe D5 that extends on $[t,\Omega_5]$, moves along the radial coordinate with $r(t)$ and has a  gauge field switched on the brane. The induced metric is
\begin{equation}
ds_{ind,D5}^2= dt^2\left( -\frac{r^2}{l^2} +\frac{l^2 \dot{r}^2}{r^2}\right) + l^2d\Omega_5^2 .   
\end{equation}
The Born-Infeld part of the action is 
\begin{equation}
S_{BI,D5}=T_{D5}~ l^5~\int dt~ d\Omega_5 \frac{r}{l}\sqrt{1-\frac{l^4 \dot{r}^2}{r^4}}= T_{D5}~l^5 \text{Vol}_{S^5}\int dt~ \frac{r}{l}\sqrt{1-\frac{l^4 \dot{r}^2}{r^4}}.\label{BID5}    
\end{equation}
In the absence of Ramond potentials $C_6=C_2=C_0=0$ and for NS $B_2=0$, but in the presence of nonzero $C_4$,  the Wess-Zumino part of the action is 
\begin{equation}
S_{WZ}=-T_{D5}\int dt ~d\Omega_5~ C_4\wedge f_2=+T_{F1}N \int a_t dt.\label{SWZD5}   
\end{equation}
Following the treatment indicated above for the case of a background generated by a stack of D3 branes with the baryon-vertex being a D5 brane probe, we find for the action of the strings
%
\begin{eqnarray}
& & S_{F1}= S_{NG} + S_{WZ-F1},\nonumber\\
& & S_{NG}=N~ T_{F1}\int dt\int_{r(t)}^{r_{UV}} dr \sqrt{-g_{tt}g_{rr}}= N T_{F1}\int dt \left[r_{UV}-r(t) \right]\;,\label{SF1}\\
& &S_{WZ-F1}=-N T_{F1} \int a_t dt.\nonumber
\end{eqnarray}
Note that these $N$ F1s cancel the charge induced by the gauge field on the compact part of the D5 brane (hence satisfying Gauss' law). In summary, the action of the system is
\begin{equation}
   S_{D5-F1}= \int dt ~\left[\alpha \sqrt{\frac{r^2}{l^2}\left(1-\frac{l^4\dot{r}^2}{r^4}\right)} +\beta \left( r_{UV}-r(t)\right)\right].\label{actionD5-baryon}  
\end{equation}
The coefficients $\alpha,\beta$ are
\begin{equation}
\alpha= T_{D5}~l^5~\text{Vol}_{S^5},~\beta= N~ T_{F1}.
\end{equation}
In particular, notice that in a static situation $r(t)=r_0$, if $\alpha=l \beta$ we have a static SUSY object. Notice that we consider the D5 probe to be non-BPS (the tension is not equal to the charge). Our probe does feel an attractive force and falls along the radial coordinate.
\\
If the coefficient $\beta=0$, we are working in a situation for which it is consistent to have  no gauge field on the brane, and no need for the F1 strings. This can occur in situations for which charge is not induced on the $\Omega_p$. In such cases the Dp branes falls following a massive geodesic. 

To continue, we study the dynamics of a composite point particle with generic action (\ref{actionD5-baryon}).
\subsection{Dynamics of the infalling baryon-vertex}\label{discuss-dynamics}
As discussed above, we are interested in the point particle Lagrangian (describing the dynamics of the composite baryon vertex) 
\begin{align}
    L=\alpha \sqrt{\frac{r^2}{l^2}\left(1-\frac{l^4\dot{r}^2}{r^4}\right)} +\beta \left( r_{UV}-r(t)\right).\label{generic-lagrangian}
\end{align}
The Euler-Lagrange equation of motion, canonical momentum and Hamiltonian are
\begin{eqnarray}
& &  \frac{d}{dt}\Big[\frac{\alpha l^3 \dot{r}}{r\sqrt{r^4-l^4\dot{r}^2}} \Big]= \beta - \frac{\alpha}{l}\times \frac{(r^4+l^4\dot{r}^2)}{r^2\sqrt{r^4-l^4\dot{r}^2}}.\label{eq-of-motionD5,D7}  \\
& &     P_r =\frac{\partial L}{\partial \dot{r}}=-\frac{\alpha l^3}{r}\frac{\dot{r}}{\sqrt{r^4 - l^4 \dot{r}^2}},\label{momentum-Pr}\\
& &
H=P_r \dot{r}-L= -\frac{\alpha r^3}{l~ \sqrt{r^4-l^4\dot{r}^2}}-\beta [r_{UV }-r(t)].\label{Hamiltonianeq}
\end{eqnarray}
Using eq.~\eqref{Hamiltonianeq}, we can invert the conserved Hamiltonian to obtain a first-order equation that solves the Euler--Lagrange equation \eqref{eq-of-motionD5,D7}. Integrating this equation gives
\begin{eqnarray}
& &
    \dot{r}=-\frac{r^2 \sqrt{l^2\Big[H +\beta (r_{UV}-r)\Big]^2-\alpha^2  r^2 }}{l^3 (H+\beta  (r_{UV}-r))}=f(r),\label{rdot}\\
    & & \int \frac{dr}{f(r)}=t \Rightarrow r(t)=- \frac{l^2 (H+\beta  r_{UV})}{\sqrt{ (H+\beta  r_{UV})^2 t^2+\alpha ^2 l^2} ~-\beta l^2}.\label{solutionr}
\end{eqnarray}
We chose the negative sign for the velocity $\dot r$ to reflect  that $r(t)$ decreases with time $t$ and also the integration constant was chosen to be $t_0=0$.

Our solution in eq.(\ref{solutionr}) describes the baryon vertex falling from $r_{UV}$ (a UV cutoff) with zero initial velocity, in agreement with the proposal of \cite{Caputa:2024sux, He:2024pox, Fan:2024iop}. First, we note that when evaluated at $t=0$, eq.(\ref{Hamiltonianeq}) implies
that the Hamiltonian has a value $H= -\frac{\alpha}{l}r_{UV}$.
We then evaluate eqs.(\ref{rdot})-(\ref{solutionr}) at $t=0$, obtaining that the initial velocity and initial position are as required.
{Expanding for $t\simeq0$, one finds
 \begin{align}
    r(t)=r_{UV}-\frac{r^3_{UV}}{2 \alpha l^4}(\alpha - \beta l)t^2 + \cdots
 \end{align}
 which clearly indicates that for $\alpha =\beta l$, the brane does not fall. In other words, the tension of F1 string balances the gravitational pull on the D5 brane (the BPS situation). On the other hand, for $\alpha > \beta l$, the non-BPS-D5 brane starts falling from $r_{UV}$ towards the interior ($r(t)<r_{UV}$) of the bulk AdS$_5$.}

Following \cite{Caputa:2024sux}, we propose
that there is a special radial coordinate $\rho$, such that when all other coordinates are constant, satisfies
$ds^2=dy^2$. This special coordinate leads to the definition of {\it proper momentum}, that is needed to calculate the complexity. In some detail,
\begin{equation}
y=\pm l~\log r.    
\end{equation}

The proposal of \cite{Caputa:2024sux}--see also
\cite{Fatemiabhari:2025cyy,Fatemiabhari:2025poq,Fatemiabhari:2026rob,
Fatemiabhari:2026goj,Li:2025fqz,Zoakos:2026obl}--relates the complexity rate to the
magnitude of the proper momentum. With the orientation-independent convention
\eqref{eq:absolute-momentum-prescription},
\begin{equation}
\dot{\mathcal C}(t)=\kappa |P_y|
=\kappa\left|\frac{\partial L}{\partial \dot r}
\frac{d\dot r}{d\dot y}\right|
=\kappa\frac{|H+\beta r_{UV}|}{l} t.
\label{eq:baryon-positive-rate}
\end{equation}
The rate is linear in time, as for an ordinary particle in Poincar\'e AdS, while the wrapped
brane and the $N$ attached F1 strings enter through the effective conserved energy. Thus
this composite object provides evidence that, within the rigid-probe approximation, the
leading quadratic growth is controlled by the collective radial motion. This is a result for
the present class of AdS configurations, not a general theorem relating boundary locality
to a universal Krylov law.

Two extensions would test rather than merely repeat the pattern found here. First,
AdS$_3\times S^3\times CY_2$ connects more directly with the AdS$_3$/CFT$_2$ setting in
which the momentum/complexity relation was originally derived and changes both the boundary
dimension and the brane realisation. The more elaborate
AdS$_3\times S^2\times CY_2\times I_\eta$ families of
\cite{Lozano:2019emq,Lozano:2019jza,Lozano:2019zvg,Lozano:2020bxo,
Lozano:2020txg} additionally introduce non-trivial flux, dilaton, source and quiver data;
they can test which of these data enter the coefficient and subleading terms of the proper
momentum. Second, AdS$_7\times S^4$ replaces the D-brane/F1 construction by an M-theory
M5/M2 realisation and changes the dual theory to the six-dimensional $(2,0)$ theory. This
would distinguish consequences of pointlikeness and AdS kinematics from features specific
to DBI actions and fundamental strings.

%
%
%

\section{Complexity measured by a falling giant-graviton}\label{complexitygiant}
We study another point-like excitation (from  the field theory point of view). We discuss the complexity calculated by a falling giant-graviton \cite{McGreevy:2000cw}. This type of solutions have been studied in various works (never applied to complexity), see for example \cite{Camino:2001at, Caldarelli:2004yk}.
We discuss the usual giant in AdS$_5\times S^5$. We use the metric and four-form potential
\begin{eqnarray}
& & ds^2=\frac{r^2}{l^2}\left(-dt^2+d\vec{x}_3^2 \right) +\frac{l^2 dr^2}{r^2}+ l^2d\Omega_5^2,\label{AdS5metrica-giant}\\
& & d\Omega_5=d\theta^2+\sin^2\theta d\phi^2+ \cos^2\theta d\Omega_3,~~4 ~d\Omega_3= \omega_1^2+\omega_2^2+\omega_3^2,~~~l^4=g_s N_c \alpha'^2,\nonumber\\
& &F_5=\frac{4}{l}(1+*) \text{Vol}_{S^5},~~C_4= \frac{r^4}{l^4} dt\wedge dx_1\wedge dx_2\wedge dx_3 -\frac{l^4}{8}\cos^4\theta d\phi\wedge \omega_1\wedge \omega_2\wedge \omega_3,\nonumber\\
& & \omega_1=\cos\psi d\alpha +\sin\psi \sin\alpha d\beta,~\omega_2=-\sin\psi d\alpha+ \cos\psi\sin\alpha d\beta,~\omega_3=d\psi+\cos\alpha d\beta.\nonumber
\end{eqnarray}
We consider a (D3-brane-) giant graviton that falls in the radial direction $r(t)$ and rotates in $\theta(t),\phi(t)$, which is a consistent configuration.

We calculate the induced metric, the Born-Infeld and the Wess-Zumino action for the giant, namely a D3 brane, that extends along $[t,S^3]$, with $r(t),\theta(t),\phi(t)$. We find
\begin{eqnarray}
& & ds^2_{ind}= -\left(\frac{r^2}{l^2} -\frac{l^2\dot{r}^2}{r^2} -l^2\dot{\theta}^2 -l^2\sin^2\theta \dot{\phi}^2\right)dt^2 +\frac{l^2}{4}\cos^2\theta \left( \omega_1^2+\omega_2^2+\omega_3^2\right),\\
& &S_{BI}=\frac{\pi^2}{4} T_{D3} l^3\int dt ~\cos^3\theta \sqrt{\frac{r^2}{l^2} -\frac{l^2\dot{r}^2}{r^2} -l^2\dot{\theta}^2 -l^2\sin^2\theta \dot{\phi}^2},\label{SBID3}\\
& & S_{WZ}=- \frac{\pi^2}{4} T_{D3}l^4 \int dt~ \cos^4\theta \dot{\phi},\nonumber\\
& & S_{BIWZ}= \nu\int dt \Bigg[\cos^3\theta \sqrt{\frac{r^2}{l^2} -\frac{l^2\dot{r}^2}{r^2} -l^2\dot{\theta}^2 -l^2\sin^2\theta \dot{\phi}^2}-\mu \cos^4\theta \dot{\phi}. \Bigg],\label{actiond3g}\\
& & \nu=\frac{\pi^2}{4}T_{D3}l^3,~~~\mu=  l.\nonumber
\end{eqnarray}
For this dynamical system, define
\begin{equation}
A=\frac{r^2}{l^2}-\frac{l^2\dot r^2}{r^2}-l^2\dot\theta^2,
\qquad
Q(\theta)=J+\mu\nu\cos^4\theta.
\label{eq:giant-AQ}
\end{equation}
The expression for the angular momentum $J=\frac{\partial L}{\partial\dot{\theta}}$ (where $L$ is the above Lagrangian), leads to
\begin{equation}
Q(\theta)=-\frac{\nu l^2\sin^2\theta\cos^3\theta\,\dot\phi}
{\sqrt{A-l^2\sin^2\theta\dot\phi^2}}.
\label{eq:giant-unsquared-charge}
\end{equation}
This fixes the physical branch that would be lost by simply squaring this relation. Solving it for $\dot{\phi}$
gives
\begin{equation}
\dot\phi=-\frac{Q(\theta)}{l\sin\theta}
\sqrt{\frac{A}{Q^2(\theta)+\nu^2l^2\sin^2\theta\cos^6\theta}}.
\label{phidot}
\end{equation}
Substitution into the partial Legendre transform yields
\begin{align}
\tilde{R}_J&= - J\dot\phi + L=-\mathcal M_J(\theta)\sqrt{A}= \mathcal M_J(\theta)
\sqrt{\frac{r^2}{l^2}-\frac{l^2\dot r^2}{r^2}-l^2\dot\theta^2}
\label{routhianmain}
\\
\mathcal M_J(\theta)&=
\frac{\sqrt{Q^2(\theta)+\nu^2l^2\sin^2\theta\cos^6\theta}}
{l\sin\theta}>0.
\label{eq:giant-corrected-Routhian}
\end{align}

This is an involved dynamical system in the $r(t),\theta(t)$ space.
We solve it below in generality. We present the solution in two steps. In the first step, we consider the possibility of setting $\theta=\theta_0$ (a constant). In the second step, we study the full $r(t)-\theta(t)$ system.
\subsection{Constant-$\theta$ reduction}
%
%
%

A constant configuration $\theta(t)=\theta_0$ is a consistent reduction only if
$\mathcal M'_J(\theta_0)=0$. Whether an interior stationary point exists depends on the
charge and brane parameters and should not be inferred from the spurious squared branch
removed above. Whenever such a stationary point exists, the reduced dynamics is
\begin{equation}
\widetilde R_J=\mathcal M_J(\theta_0)
\sqrt{\frac{r^2}{l^2}-\frac{l^2\dot r^2}{r^2}},
\label{eq:giant-constant-theta-corrected}
\end{equation}
which is the action of a particle of positive effective mass
$\mathcal M_J(\theta_0)$ in AdS$_5$. Its fixed-charge proper momentum therefore gives
$\mathcal C_J(t)-\mathcal C_J(0)\propto t^2$. We shall not assume that this special branch
exists for generic parameters.

Let us now study the complete system with $r(t)$ and $\theta(t)$.
\subsection{The complete system $r(t),\theta(t)$}\label{giantfull}

We use the positive reduced Routhian in eq.~\eqref{routhianmain}. The momentum conjugate
to $\theta$, its equation of motion and the conserved reduced energy are
\begin{align}
P_\theta&=\frac{\partial\tilde{R}_J}{\partial\dot{\theta}}=-\frac{l^2\mathcal M_J(\theta)\dot\theta}
{\sqrt{\frac{r^2}{l^2}-\frac{l^2\dot r^2}{r^2}-l^2\dot\theta^2}},\nonumber\\
\frac{d}{dt}\left(\frac{\partial \tilde{R}_J}{\partial \dot{\theta}}\right)&=\frac{\partial\tilde{R}_J}{\partial\theta}\longrightarrow \dot P_\theta=\mathcal M'_J(\theta)
\sqrt{\frac{r^2}{l^2}-\frac{l^2\dot r^2}{r^2}-l^2\dot\theta^2},\nonumber\\
\mathcal E_J&=\widetilde R_J-P_r\dot r-P_\theta\dot\theta
=\frac{\mathcal M_J(\theta)r^2}
{l^2\sqrt{\frac{r^2}{l^2}-\frac{l^2\dot r^2}{r^2}-l^2\dot\theta^2}}>0.
\label{hamilrouth}
\end{align}
A second conserved quantity is
\begin{equation}
K=P_\theta^2+l^2\mathcal M_J^2(\theta),
\qquad \dot K=0.
\label{conservedK}
\end{equation}
The two constants $(\mathcal E_J,K)$ give
\begin{align}
\dot r&=\pm\frac{r^2}{\mathcal E_Jl^4}
\sqrt{\mathcal E_J^2l^4-Kr^2},\nonumber\\
r(t)&=\frac{\mathcal E_Jl^2}
{\sqrt{K+\mathcal E_J^2(t-t_0)^2}},
\label{rdetgiant}\\
l\dot\theta&=\pm
\sqrt{1-\frac{l^2\mathcal M_J^2(\theta)}{K}}
\sqrt{\frac{r^2}{l^2}-\frac{l^2\dot r^2}{r^2}},
\label{angelito}\\
\int_{\theta_0}^{\theta(t)}
\frac{d\vartheta}{\sqrt{1-l^2\mathcal M_J^2(\vartheta)/K}}
&=\pm\arctan\left[\frac{\mathcal E_J}{\sqrt K}(t-t_0)\right].
\label{thetadetgiant}
\end{align}
Together with the physical angular velocity in eq.~\eqref{phidot}, these equations give the
full solution, implicitly in $\theta(t)$, of the original system in
eq.~\eqref{actiond3g}. The radial coordinate retains the particle-like form, while the
charge and the wrapped-brane dynamics enter through $\mathcal M_J(\theta)$ and $K$.

For the comparison with spread complexity we choose the reference time $t_0=0$ and impose
rest in the complete reduced configuration space,
\begin{equation}
r(0)=r_{UV},\qquad \dot r(0)=0,\qquad
\theta(0)=\theta_0,\qquad \dot\theta(0)=0.
\label{eq:giant-reduced-rest-data}
\end{equation}
These conditions fix
\begin{equation}
K=l^2\mathcal M_J^2(\theta_0),
\qquad
\mathcal E_J=\frac{\mathcal M_J(\theta_0)r_{UV}}{l}.
\label{eq:giant-rest-constants}
\end{equation}
They are the bulk counterpart of choosing the state at the reference time as the initial
Krylov seed. Initial data with $\dot\theta(0)\neq0$ describe a different bulk preparation
and are not identified here with the standard Krylov construction about $t=0$.

Let us take a small detour to write the units of all the quantities that appear in this section. We find
\begin{eqnarray}
& &[r]=[x_i]=[l]=\text{length}, ~[\theta]=[\phi]=[\omega_i]=1. ~[\text{Lagrangian and Routhian}]=\frac{1}{\text{length}}, \nonumber\\
& & [T_{D3}]=\frac{1}{\text{length}^4},[\mu]=\text{length},~[\mathcal M_J(\theta)]=[\nu]=[\mathcal E_J]=\frac{1}{\text{length}}, ~[P_\theta]=[K]=1.\label{unitsgiant}
\end{eqnarray}
With these assignments of units, all the equations in this section are dimensionally consistent.
\\
Let us  calculate the complexity associated with the falling giant graviton.
\subsection{Complexity of the falling giant}\label{giantfullc}

The coordinate $\phi$ was eliminated at fixed $J$ in constructing the Routhian and must
not be restored as an independent direction in the proper-distance element. The reduced
configuration space is parametrised by $(r,\theta)$, with
\begin{equation}
\dot y_J^{2}=\frac{l^2\dot r^2}{r^2}+l^2\dot\theta^2,
\qquad
\widetilde R_J=\mathcal M_J(\theta)
\sqrt{\frac{r^2}{l^2}-\dot y_J^{2}}.
\label{proper-coordinate-giant}
\end{equation}
The fixed-charge proper momentum is therefore computed from the same Routhian that
generates the dynamics,
\begin{equation}
P_y^{(J)}=\frac{\partial\widetilde R_J}{\partial\dot y_J}
=-\frac{\mathcal M_J(\theta)\dot y_J}
{\sqrt{r^2/l^2-\dot y_J^{,2}}}
=-\frac{\mathcal E_Jl^2}{r^2}\dot y_J.
\label{propermom-explicit}
\end{equation}
Note that  {\it there is no separate $P_\phi\partial\dot\phi/\partial\dot y_J$ term and no explicit
$l^2\sin^2\theta\dot\phi^2$ in $\dot y_J^2$}. The Wess--Zumino coupling has not been lost:
it shifts the fixed charge through $Q(\theta)=J+\mu\nu\cos^4\theta$ inside
$\mathcal M_J(\theta)$. This is precisely the compatibility required between the
fixed-charge dynamics and the complexity prescription.

Using eq.~\eqref{eq:absolute-momentum-prescription}, we identify
\begin{equation}
\dot{\mathcal C}_J(t)=\kappa|P_y^{(J)}(t)|.
\label{enzof}
\end{equation}
For the reduced rest data \eqref{eq:giant-reduced-rest-data}, the equations of motion give
\begin{align}
\dot r(t)&=-\frac{r_{UV}^3}{l^4}t+\mathcal O(t^3),\nonumber\\
\dot\theta(t)&=-\frac{r_{UV}^2}{l^4}
\frac{\mathcal M'_J(\theta_0)}{\mathcal M_J(\theta_0)}t
+\mathcal O(t^3),\nonumber\\
|P_y^{(J)}(t)|&=
\frac{\mathcal M_J(\theta_0)r_{UV}}{l^2}
\sqrt{1+\left[\frac{\mathcal M'_J(\theta_0)}
{\mathcal M_J(\theta_0)}\right]^2}\,|t|+\mathcal O(|t|^3).
\label{eq:giant-fixedJ-short-time}
\end{align}
Consequently $\dot{\mathcal C}_J(0)=0$ and
$\mathcal C_J(t)-\mathcal C_J(0)=\mathcal O(t^2)$. On a constant-$\theta$ branch the
coefficient reduces to $\mathcal E_J/l$. Appendix \ref{appendixc} supplies the details of
this expansion.

The fixed-charge late-time behaviour can also be read analytically. Provided
$\mathcal M_J(\theta(t))$ remains finite along the asymptotic trajectory,
eqs.~\eqref{rdetgiant} and \eqref{propermom-explicit} imply
\begin{equation}
r(t)=\frac{l^2}{|t|}+\mathcal O(|t|^{-3}),
\qquad
\dot y_J=\frac{l}{|t|}+\mathcal O(|t|^{-3}),
\qquad
|P_y^{(J)}|=\frac{\mathcal E_J}{l}|t|+\mathcal O(|t|^{-1}).
\label{eq:giant-fixedJ-late-time}
\end{equation}
Thus the quadratic leading growth follows from the reduced proper momentum without
reintroducing the cyclic motion. The analytic expressions above are the quantities
relevant to a definite-$J$ sector.

Let us now study the falling dynamics of an extended probe, proposed as a collective model for the complexity of a non-local excitation.

\section{Spread complexity for a non-local  (extended) operator}\label{complexityextended-string}

The fixed-charge particle of section \ref{CONSRSR}, the baryon vertex of section
\ref{sec:BV} and the giant graviton of section \ref{complexitygiant} share a linear
late-time proper-momentum rate within the rigid Poincar\'e-AdS configurations analysed here.
This common behaviour is controlled by collective radial infall, while charge and brane
structure enter the coefficient and subleading terms. This behaviour should not be read as a theorem
that every boundary-local probe has $\mathcal C\sim t^2$ for large times. In fact, the examples mentioned share a highly
symmetric, zero-temperature conformal background and a collective-coordinate truncation.

Instead, in this section, we explore a different aspect of the above proposal, namely we consider the spread complexity of an {\it non-local (extended) operator}.
We choose a string that falls in AdS under gravity and is extended along one spatial direction. This is the holographic version of a line operator. We also {\it adopt} the proposal that the rate-of-change of the Krylov complexity $\dot{C}(t)$ is proportional to the proper momentum \cite{Caputa:2024sux, He:2024pox, Fan:2024iop}.

To calculate its spread complexity, we consider an F1  string
parametrised by the $(\sigma,\tau)$ coordinates, that moves according to
\begin{equation}
   t=\tau, ~~x=x(\sigma),~~r=r(\tau). \label{embeddingstring}
\end{equation}
Strictly speaking, one should begin with a generic embedding
\begin{equation*}
   t=t(\sigma,\tau),~ r=r(\sigma,\tau), ~x=x(\sigma,\tau). 
\end{equation*} 
Then, one studies the corresponding equations of motion, and shows that the embedding in eq.\eqref{embeddingstring} is a consistent truncation. On the embedding (\ref{embeddingstring}), the action of the massive probe reads
\begin{eqnarray}
& &  S= \int dt L~~;~~L=-m \sqrt{\frac{r^4}{l^4}-\dot{r}^2}.\label{accioncuerda}\\
& & 
\text{We defined}~~ m= \frac{L_x}{2\pi \alpha'}~~\text{ and }~\int d\sigma ~x'(\sigma)=L_x.\nonumber
\end{eqnarray}
The equation of motion and conserved Hamiltonian are
\begin{eqnarray}
& & -\frac{d}{dt}\left[\frac{\dot{r}}{\sqrt{\frac{r^4}{l^4} -\dot{r}^2}} \right]= \frac{2 r^3}{l^4\sqrt{\frac{r^4}{l^4} -\dot{r}^2}},\label{eomstrings}\\
& &H= m\frac{r^4}{l^4\sqrt{ \frac{r^4}{l^4} -\dot{r}^2  }},~~~\Rightarrow \dot{r}(t)=-\frac{r^2}{Hl^4}\sqrt{H^2 l^4 - m^2 r^4}.\label{cantidades}\\
& & \nonumber
\end{eqnarray}
Finding $\ddot{r}$ by derivative of $\dot{r}$ and writing everything in terms of $r(t)$ and  $\dot{r}(t)$, the equation of motion in (\ref{eomstrings}) is satisfied by the first order equation in (\ref{cantidades}). Furthermore, we choose a string that starts falling from a  position $r_{UV}$ with zero initial velocity. This condition, together with eq.(\ref{cantidades}) imply that 
\begin{equation}
r_{UV}^4=\frac{H^2 l^4}{m^2}.\label{ruvstrings}    
\end{equation}
Integrating eq.\eqref{cantidades}, we find
\begin{eqnarray}
& & t(r)=\frac{l^2}{r}\Bigg[{}_2F_1\left(-\frac14,\frac12,\frac34,\frac{r^4}{r_{UV}^4} \right)- \frac{r}{r_{UV}}\sqrt{\pi}\,\frac{\Gamma(\frac{3}{4})}{\Gamma(\frac{1}{4})}\Bigg]. \label{solutionstring}
%
%
\end{eqnarray}
The integration constant was chosen to satisfy $t(r_{UV})=0$ or conversely $r(0)=r_{UV}$.
\begin{figure}
    \centering
    \includegraphics[width=0.7\linewidth]{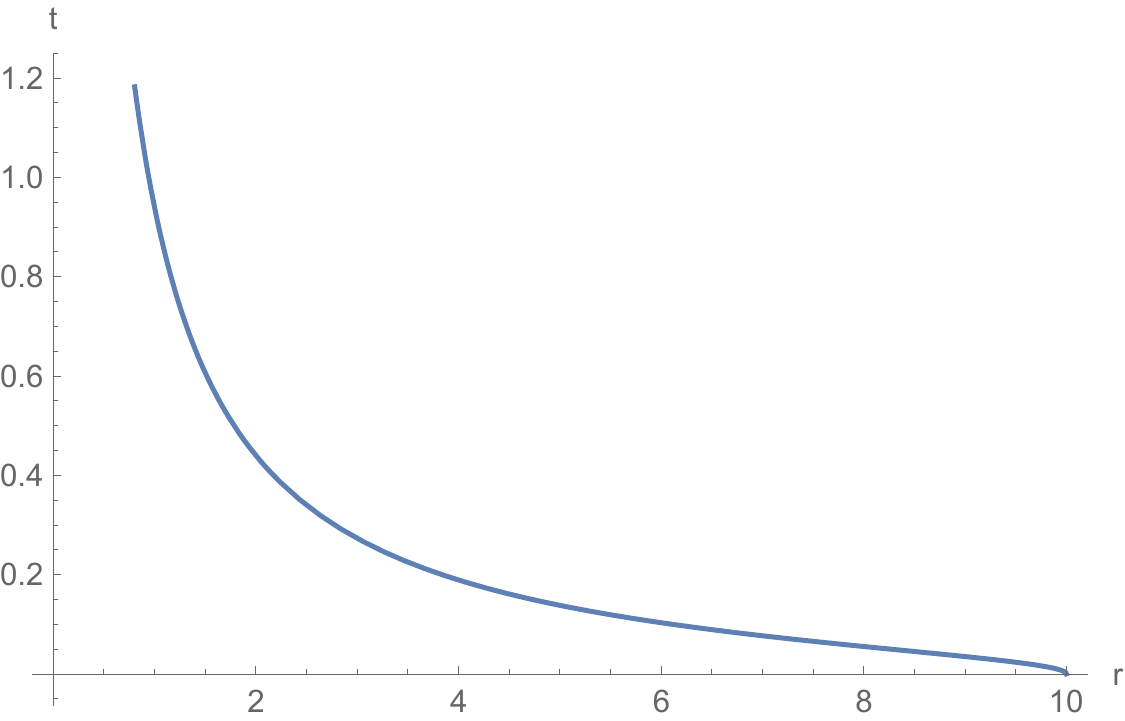}
    \caption{We plot $t$ vs $r$, where we set $l=1$ and $r_{UV}=10$.}
    \label{figtr}
\end{figure}
\\
In Figure \ref{figtr} we give the plot of $t(r)$. This shows that the string takes infinite time to reach $r=0$.
\\
Expanding the expression (\ref{solutionstring})  close to $r=r_{UV}$ gives
%
\begin{align}
    \frac{r_{UV}}{l^2}t\approx\sqrt{1-\frac{r}{r_{UV}}} + {\cal O}\left(\left(1-\frac{r}{r_{UV}}\right)^{\frac32}\right).
\end{align}
%
Inverting this yields for short times
\begin{align}
\label{e7.17}
   r(t)\approx r_{UV}\Bigg[1-\frac{r^2_{UV}}{l^4}t^2\Bigg],~~ \text{for}~t\to 0.
\end{align}
This satisfies the conditions that the string starts at $r_{UV}$ for $t=0$ and has a zero initial velocity (along the radial direction), $\dot{r}|_{t=0}.$

On the other hand, expanding eq.\eqref{solutionstring} near $\frac{r}{r_{UV}} \sim 0^+$ gives
\begin{eqnarray}
& &\frac{r_{UV}}{l^2}(t+ t_0)\approx  \frac{r_{UV}}{r} -\frac{1}{6}\left(\frac{r}{r_{UV}}\right)^3 -\frac{3}{56}\left( \frac{r}{r_{UV}}\right)^7+.... \;,\\
& &\text{where}~~t_0= \frac{l^2}{r_{UV}}\frac{\sqrt{\pi} \Gamma(\frac{3}{4})}{\Gamma(\frac14)}.\nonumber
\end{eqnarray}
%

\begin{figure}
    \centering
    \includegraphics[width=0.6\linewidth]{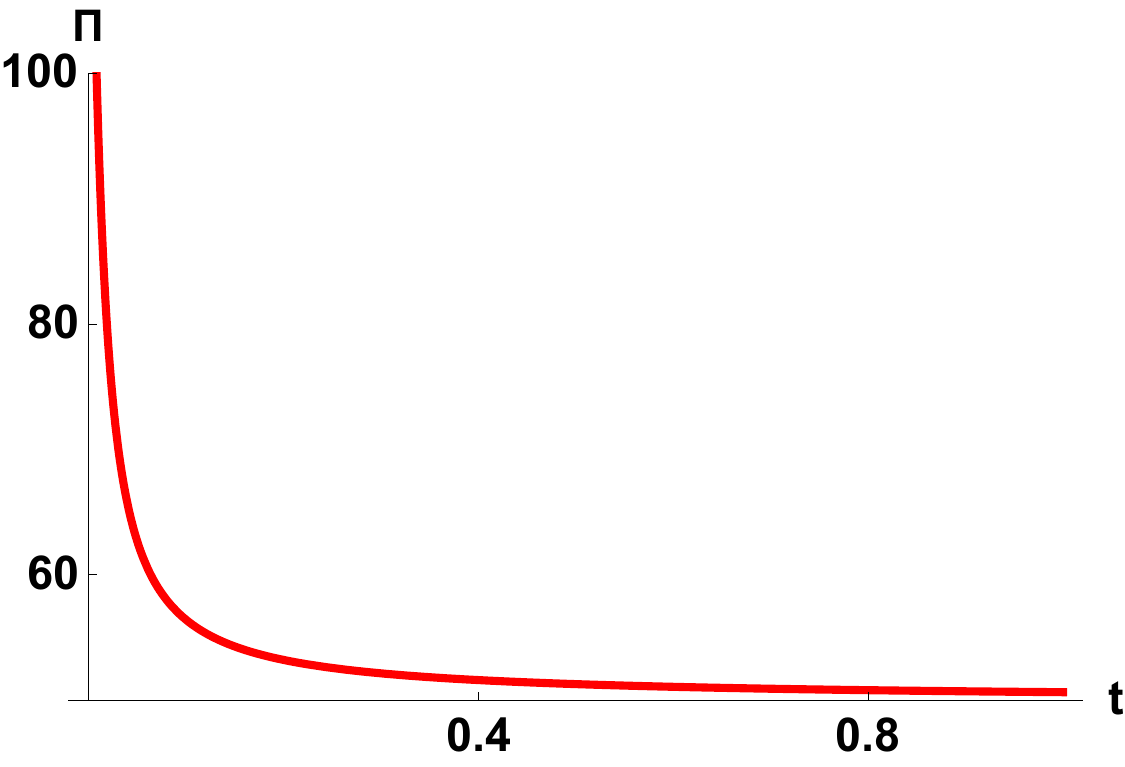}
    \caption{ $\Pi$ vs. $t$ plot. We set $m=0.001$, $l=1$,  $r_{UV}=50$.}
    \label{figPi}
\end{figure}
\begin{figure}
    \centering
    \includegraphics[width=0.6\linewidth]{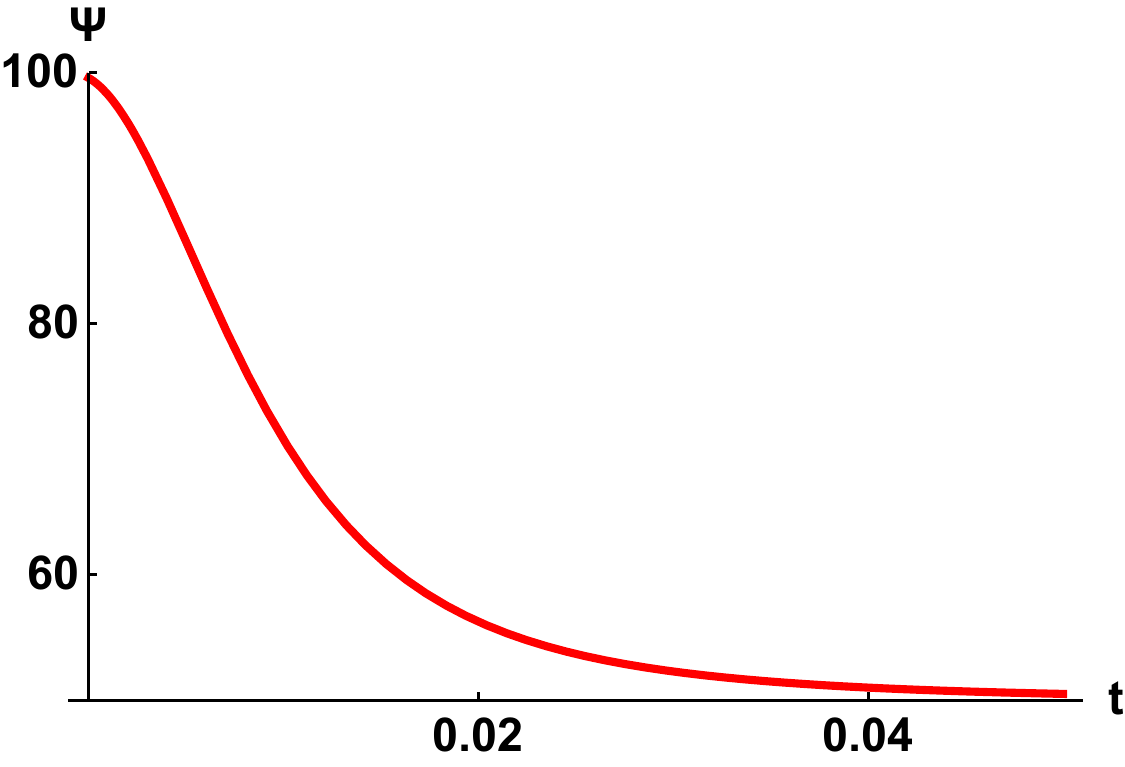}
    \caption{ $\Psi$ vs. $t$ plot. We set $m=0.001$, $l=1$,  $r_{UV}=50$.}
    \label{figpsi}
\end{figure}
%
%
This is inverted to obtain the result that for large times $r(t)$ behaves as
\begin{align}
\label{e7.20}
    r(t)=\frac{l^2}{(t+{t}_0)} -\frac{l^{10}}{6 r_{UV}^4 (t+t_0)^5}+ {\cal O}\left(\frac{1}{(t+t_0)^9}\right)~;~~~t \gg {t}_0.
\end{align}
\\
\\
\underline{\bf The Holographic Krylov Complexity}
\\
\\
To calculate the complexity, we adopt the working hypothesis that this is proportional to the proper momentum of the falling string, extending the proposal of \cite{Caputa:2024sux, Fan:2024iop, He:2024pox}.  We change to the proper coordinate as proposed in \cite{Caputa:2024sux, Fatemiabhari:2026goj, Fatemiabhari:2025poq, Fatemiabhari:2026rob},
\begin{equation}
ds^2=dy^2= \frac{l^2}{r^2}dr^2~~\to~~ y=\int \frac{l}{r}dr=l~ \log r   .
\end{equation}
\begin{equation}
\dot{\mathcal C}=\kappa|P_y|
=\kappa\left|P_r\frac{d\dot r}{d\dot y}\right|
=\kappa\frac{ml|\dot r|}{r\sqrt{1-l^4\dot r^2/r^4}}.
\label{complexityF1}
\end{equation}
For {\it short times}, the rate of change of the complexity is linear in time, as is conventionally observed in CFTs \cite{Caputa:2024sux}. Indeed, using \eqref{e7.17}, the  early time rate-of-change of the complexity is
\begin{align}
\dot{\mathcal C}(t)\approx\kappa\left[\frac{2m r_{UV}^2}{l^3}t
+\frac{6m r_{UV}^4}{l^7}t^3+\mathcal O(t^5)\right],
\qquad t\to0.
\label{compstringt=0}
\end{align}
%
On the other hand, for {\it large times}, we use the expansion in eq.(\ref{e7.20}) and find
\begin{equation}
\dot{\mathcal C}(t)\approx\kappa\left[
\frac{m r_{UV}^2}{l^3}(t+t_0)
-\frac{ml^5}{3r_{UV}^2(t+t_0)^3}
+\mathcal O(t^{-4})\right],
\qquad t\to\infty.
\label{complexitylargetimestring}
\end{equation}
Hence, for large times, the rate-of-change of the complexity is also linear at leading order.
\\
In summary, we find that for both short and large times, the rate of change of the complexity is to leading order, linear in time. This is the same behaviour as observed for point-like, uncharged and structureless probes of mass $m$ (dual to local operators of conformal dimension $m$) \cite{Caputa:2024sux, Fan:2024iop, He:2024pox}  
\begin{equation}
\dot{\mathcal C}_{\text{particle}}(t)
=\kappa\frac{|H|}{l}t
=\kappa\frac{m r_{UV}}{l^2}t,
\qquad t>0.
\label{complexityparticle}
\end{equation}
What is new and interesting for the non-local operators (dual to extended probes in gravity) are the subleading terms and the intermediate regimes; details that are absent in the case of a massive point-like probe of eq.(\ref{complexityparticle}). In Figure \ref{figPi}, we plot the quotient 
\begin{equation}
\Pi=\frac{\text{rate of change for complexity of string}}{\text{rate of change for complexity of particle}}=\frac{\dot{C}_{\text{string}}}{\dot{C}_{\text{particle}}}=\frac{m\sqrt{r_{UV}^4- r(t)^4}}{ H~ r(t)}\frac{1}{t}.    
\end{equation}
%
{Consider a particle and a string that fall from the same position $r_{UV}$ and have the same mass $m$. Using the  early time expansion for $\dot{C}_{\text{string}}$ in \eqref{compstringt=0}, one finds
\begin{align}
\label{e5.16}
    \Pi|_{t \sim 0}= \frac{2r_{UV}}{l}
\end{align}
which is the (large) constant depicted in Figure \ref{figPi}.}
{On the other hand, using the late time expansion in \eqref{complexitylargetimestring}, one finds
\begin{align}
     \Pi|_{t \gg 1}= \frac{r_{UV}}{l},
\end{align}
 also reflected in Figure \ref{figPi}.}
To refine the comparison, we define the quotient
\begin{equation}
\Psi=\frac{\text{second derivative of complexity of string}}{\text{second derivative of complexity of particle}}=\frac{\ddot{C}_{\text{string}}}{\ddot{C}_{\text{particle}}}=\frac{r_{UV}^4 +r^4(t)}{l~r_{UV}^3}.
\end{equation}
{Close to the UV-cutoff in the geometry this yields $\Psi = \frac{2r_{UV}}{l}$, which is same as in \eqref{e5.16}. On the other hand, at late times, it is easy to see that $\Psi =\frac{r_{UV}}{l}$, see Figure \ref{figpsi}.}
\\
With this, we hope to have made the point that extended operators display a different complexity than point-like operators do. These differences are apparent in the short and long times subleading behaviour. It would be interesting to study the effect of string fluctuations on the complexity. Of course, having a purely field theoretical calculation showing similar behaviour would be very valuable and we leave this for future investigations. It might be possible to study our string in the context of \cite{Das:2024tnw}.
Let us present some conclusions and future directions.

\section{Conclusions and Future Directions}\label{section-concl}

In this work we tested the proper-momentum proposal for holographic spread complexity using
probes with conserved charge, composite brane structure and genuine spatial extension. The
analysis exposes an important distinction. If an internal coordinate is cyclic, its Noether
charge must be fixed before the proper momentum is defined. The corresponding Routhian
retains the physical charge dependence while removing the spurious constant initial rate
produced by counting the conjugate velocity as a Krylov-spreading direction. Both the
$R$-charged particle and the giant graviton then satisfy the universal quadratic short-time
onset of unitary spread complexity. The unreduced momentum remains a useful diagnostic of
the complete bulk trajectory, but it is not the complexity rate in a sector of definite
charge.
\\
The baryon vertex and the spatially stretched fundamental string probe different kinds of
structure. In the former, wrapped-brane dynamics and attached strings modify the effective
energy of an object that remains pointlike to the boundary observer. In the latter, the
non-local extension changes the radial effective action and produces distinct subleading and
intermediate-time behaviour. Across the rigid Poincar\'e-AdS configurations studied here,
the fixed-charge proper momentum is linear at late times. This common scaling is evidence
for a robust pattern in the tested class, but the present examples do not establish a
universality theorem. In particular, no general identification can be made between the
gauge-group rank and the number of sites of a Krylov chain: the Krylov dimension depends on
the seed, the symmetry sector, degeneracies and the effective Hilbert space.
\\
The most direct next step is a boundary calculation. One should project a concrete charged
state or operator onto a definite symmetry sector, determine its survival amplitude and
sectoral Lanczos coefficients, and compare the resulting spread complexity with the bulk
Routhian \cite{Caputa:2025mii,Caputa:2025ozd}. More precisely, the open problem is to
identify a controlled large-$N$, fixed-charge regime of the boundary theory in which this
comparison can be carried out. It would also be
valuable to include worldsheet and worldvolume fluctuations, thereby testing the limits of
the rigid collective-coordinate approximation.
\\
Finally, backgrounds that vary the spacetime dimension, flux realisation and infrared
geometry provide targeted tests of the interpretation. AdS$_3\times S^3\times CY_2$ is
closer to the original AdS$_3$/CFT$_2$ original derivation. AdS$_7\times S^4$ replaces DBI/F1
physics by an M5/M2 construction. Confining, thermal and black-hole backgrounds can
test whether an infrared scale or horizon changes the separation between collective radial
growth and probe-dependent data. The families in
\cite{Anabalon:2021tua,Nunez:2023nnl,Nunez:2023xgl,Chatzis:2024kdu,
Chatzis:2024top,Chatzis:2025hek,Anabalon:2023lnk,Anabalon:2024che,
Anabalon:2026yxk,Fatemiabhari:2025usn} offer concrete laboratories. These extensions will
determine which features follow from AdS kinematics and which retain detailed information
about charge, compositeness or spatial extension.

\section*{Acknowledgments}
For discussions, for comments on the manuscript, and for sharing their ideas with us, we wish to thank: Pawel Caputa, Dimitrios Chatzis, Ali Fatemiabhari,  Madison Hammond, Yolanda Lozano, Alfonso Ramallo, Diego Rodriguez-Gomez, Ricardo Santamaria. C. N. is supported by STFC’s grants ST/Y509644- 1, ST/X000648/1 and ST/T000813/1. D.R. would like to acknowledge the Mathematical Research Impact Centric Support (MATRICS) grant (MTR/2023/000005) received from ANRF, India.
The work of HN is supported in part by  CNPq grant 
304583/2023-5 and FAPESP grant 2019/21281-4.
HN would also like to thank the ICTP-SAIFR for their support 
through FAPESP grant 2021/14335-0, and to CEA-Saclay for their 
hospitality during a part of this project.

 \appendix

\section{The baryonic vertex in case of AdS$_3\times S^3\times CY_2$ }\label{appendixb}
We study the complexity of the D1-D5 two dimensional SCFT measured by a composite operator of the baryon-vertex kind. We probe the spacetime with a D7 brane that extends along $[t,S^3,\text{CY}_2] $ and falls along a geodesic parametrised by $r(t)$. The metric of the spacetime  is given by
\begin{align}
    ds^2=\frac{r^2}{l^2}\left(-dt^2+d\vec{x}^2 \right)+\frac{l^2 dr^2}{r^2} +\mu^2~ d\Omega^2_3 +\nu^2 ~ds^2_{CY_2}.
\end{align}
The parameters $(\mu,\nu)$ indicate the radii of $S^3$ and the Calabi-Yau manifold. These are related to the numbers $(N_1,N_5)$ of D1 and D5 branes via quantisation of fluxes. A pure-gauge one-form is switched on the probe D7 brane. We  consider the electric $F_3$ in this configuration or the magnetic $F_7$ that indicates the presence of D1 branes.
The induced metric on the D7 probe is 
\begin{equation}
ds_{ind,D7}^2= \frac{r^2}{l^2}\left(-1+\frac{l^4 \dot{r}^2}{r^4}\right) dt^2+\mu^2~ d\Omega^2_3 +\nu^2 ~ds^2_{CY_2}.\label{D7-induced}   
\end{equation}
The Born-Infeld part of the action of this D7 brane is
\begin{equation}
 S_{BI, D7}=T_{D7}\mu^3 \nu^4 \int dt  ~\sqrt{\frac{r^2}{l^2}\left(1-\frac{l^4 \dot{r}^2}{r^4}\right)}.\label{SBID7} 
\end{equation}
The Wess-Zumino term on the D7 consists only of the contribution from the $C_6$ background field (with $F_7=dC_6$). In the absence of other background fluxes, and switching on a gauge field on the D7 probe, we find that for this configuration
\begin{eqnarray}
& & S_{WZ, D7}= -T_{D7}\int_{t\times S^3\times CY_2} C_6 \wedge f_2=T_{F1}N_{D1}\int dt ~a_t,\label{SWZD7} \\
& & N_{D1}= \int_{S^3\times \text{CY}_2} F_7.\nonumber
\end{eqnarray}
The WZ term is canceled by the presence of $N_{D1}$ F1 strings that extend from the D7 probe and a D-brane fixed at $r_{UV}$. As above, these strings are modeled here as being 'rigid' having no kinetic term, but contributing to the action of the composite object with a potential term. As is characteristic of strings, we model this contribution as being proportional to the length of the string. This gives
\begin{equation}
    S_{F1}= N_{D1} ~T_{F1}\int dt~\left[r_{UV}- r(t)\right].
\end{equation}
As a result the full Lagrangian for the system is of the same form as in eq.(\ref{actionD5-baryon}), with the coefficients $\alpha=T_{D7}\mu^3\nu^4 \text{Vol}_{S^3\times CY_2}$ and $\beta= T_{F1} N_{D1}$. This dynamics is studied in Section \ref{discuss-dynamics}. One can make a Routhian analysis, like the one described in Appendix \ref{appendixa} below.

 \section{Baryon vertex using the Routhian}\label{appendixa}
Consider the D5 brane probe on AdS$_5\times S^5$ in Section \ref{sec:BVADS5}. 
The metric and Ramond four form potential are
\begin{eqnarray}
& & ds^2= \frac{r^2}{l^2}\left(-dt^2+ d\vec{x}_{3}^2 \right) +\frac{l^2 dr^2}{r^2} + l^2d\theta^2+l^2\sin^2\theta d\Omega_4,\label{metrciaads5xs5appe}  \\
& & C_4=4l^4\left( \frac{3\theta}{8} -\frac{\sin2\theta}{4}+ \frac{\sin4\theta}{32}\right)\text{Vol}_{\Omega_4} = {\cal C}_4(\theta) \text{Vol}_{\Omega_4},\nonumber\\
& & F_4=dC_4=4l^4 \sin^4\theta~ d\theta \wedge \text{Vol}_{\Omega_4}.\nonumber
\end{eqnarray}
We  probe this spacetime with a D5 brane extended along the five sphere and time. We switch on a gauge field on the brane $a_1=a_t(\theta) dt$, with nonzero curvature $f_2= a_t'(\theta)d\theta\wedge dt= f_{\theta t}d\theta\wedge dt$. As in the main body of the paper, we consider a brane embedding of the type
\begin{equation}
[t,\Omega_5],~~~r(t).
\end{equation}
The Born Infeld and the Wess-Zumino contributions to the probe brane action are
\begin{eqnarray}
& & S_{BI}=l^5T_{D5}\int dt ~d\theta~ d\Omega_4 ~\sin^4\theta ~\sqrt{\frac{r^2}{l^2}\left(1-\frac{l^4 \dot{r}^2} {r^4} \right)- f_{\theta t}^2},\nonumber\\
& &=l^5 T_{D5}\text{Vol}_{\Omega_4} \int dt d\theta \sin^4\theta ~\sqrt{\frac{r^2}{l^2}\left(1-\frac{l^4 \dot{r}^2} {r^4} \right)- f_{\theta t}^2}, ~~~  \label{BIA}\\
& & S_{WZ}=-T_{D5}\int_{\Omega_4,\theta,t}C_4\wedge f_2=
-T_{D5}\text{Vol}_{\Omega_4}\int dt d\theta {\cal C}_4(\theta) f_{\theta t},\nonumber\\
& &S_{BIWZ}= T_{D5}\text{Vol}_{\Omega_4}\int dt d\theta \Bigg[l^5\sin^4\theta ~\sqrt{\frac{r^2}{l^2}\left(1-\frac{l^4 \dot{r}^2} {r^4} \right)- f_{\theta t}^2}- {\cal C}_4(\theta)f_{\theta t}\Bigg].
\end{eqnarray}
We denote $\mu=T_{D5}\text{Vol}_{\Omega_4}$  and define the conjugate momentum 
\begin{equation}
    \Pi_\theta=\frac{\delta S_{BIWZ}}{\delta f_{\theta t}}=\mu\left( -\frac{l^5f_{\theta t}\sin^4\theta}{\sqrt{\frac{r^2}{l^2}\left(1-\frac{l^4 \dot{r}^2} {r^4} \right)- f_{\theta t}^2} }-{\cal C}_4(\theta)\right).\label{momentumRouthian}
\end{equation}
We can find $f_{\theta t}$ in terms of $\Pi_\theta$,
\begin{equation}
 f_{\theta t}= \pm (\Pi_{\theta} +\mu {\cal C}_4) \sqrt{\frac{\frac{r^2}{l^2}(1-\frac{l^4 \dot{r}^2} {r^4})}{(\Pi_\theta +\mu {\cal C}_4)^2+ \mu^2 l^{10}\sin^8\theta}} . 
\end{equation}
Then, we define the Routhian, a Legendre transform of the Lagrangian along the 'cyclic coordinate' $f_{\theta t}$. We have
\begin{eqnarray}
& &  R=-L +\Pi_\theta f_{\theta t}   ,\label{routhian}\\
& & R= \int d\theta \sqrt{\left(\Pi_\theta +\mu {\cal C}_4(\theta) \right)^2 +\mu^2 l^{10}\sin^8\theta} \times \int dt ~\sqrt{\frac{r^2}{l^2}\left(1-\frac{l^4 \dot{r}^2} {r^4} \right)},\nonumber\\
& & R=\alpha \times \int dt ~\sqrt{\frac{r^2}{l^2}\left(1-\frac{l^4 \dot{r}^2} {r^4} \right)}.\nonumber
\end{eqnarray}
Using the Routhian, we find the same dynamics that we study in eq. (\ref{actionD5-baryon}). 

To obtain the contribution coming from Gauss' law, we observe that the momentum $\Pi_\theta$ is piece-wise constant away from the sources. In other words $\Pi_\theta$ plays the role of the electric displacement $\vec{D}$-field. That is, \begin{equation}
\vec{\nabla}\cdot\vec{D}=\partial_\theta \Pi_\theta=\sum_{i}q_i \delta(\theta-\theta_i). 
\end{equation}
Using this equation, we conclude that Gauss' law on the compact $\theta$-space requires F1 strings that carry away the induced flux. We must therefore add the F1 action in eq.~\eqref{SF1}. This gives the same dynamical contribution as the treatment in the main text.
\section{Fixed-charge expansion for the falling giant graviton}\label{appendixc}

This appendix gives the short-time expansion appropriate to the fixed-$J$ prescription of
sections \ref{giantfull} and \ref{giantfullc}. The cyclic angle $\phi$ has already been
eliminated, and the initial data are at rest in the reduced $(r,\theta)$ configuration space,
as in eq.~\eqref{eq:giant-reduced-rest-data}. Consequently
\begin{equation}
K=l^2\mathcal M_J^2(\theta_0),
\qquad
\mathcal E_J=\frac{\mathcal M_J(\theta_0)r_{UV}}{l}.
\label{eq:appendixC-rest-constants}
\end{equation}
Expanding the radial solution \eqref{rdetgiant} gives
\begin{equation}
r(t)=r_{UV}\left[1-\frac{r_{UV}^2}{2l^4}t^2
+\frac{3r_{UV}^4}{8l^8}t^4+\mathcal O(t^6)\right],
\qquad
\dot r(t)=-\frac{r_{UV}^3}{l^4}t+\mathcal O(t^3).
\label{eq:appendixC-r-expansion}
\end{equation}
The $\theta$ equation follows either from the reduced Euler--Lagrange equation or by
expanding eq.~\eqref{thetadetgiant} about the endpoint at which
$K=l^2\mathcal M_J^2(\theta_0)$. One finds
\begin{equation}
\theta(t)=\theta_0-\frac{r_{UV}^2}{2l^4}
\frac{\mathcal M'_J(\theta_0)}{\mathcal M_J(\theta_0)}t^2
+\mathcal O(t^4),
\qquad
\dot\theta(t)=-\frac{r_{UV}^2}{l^4}
\frac{\mathcal M'_J(\theta_0)}{\mathcal M_J(\theta_0)}t
+\mathcal O(t^3).
\label{eq:appendixC-theta-expansion}
\end{equation}
It follows that the reduced proper velocity is
\begin{equation}
|\dot y_J(t)|=\frac{r_{UV}^2}{l^3}
\sqrt{1+\left[\frac{\mathcal M'_J(\theta_0)}
{\mathcal M_J(\theta_0)}\right]^2}\,|t|+\mathcal O(t^3).
\label{eq:appendixC-y-expansion}
\end{equation}
Substitution in eq.~\eqref{propermom-explicit} yields
\begin{equation}
|P_y^{(J)}(t)|=
\frac{\mathcal M_J(\theta_0)r_{UV}}{l^2}
\sqrt{1+\left[\frac{\mathcal M'_J(\theta_0)}
{\mathcal M_J(\theta_0)}\right]^2}\,|t|+\mathcal O(t^2).
\label{eq:appendixC-Py-expansion}
\end{equation}
Therefore $\dot{\mathcal C}_J(0)=0$ and the fixed-charge complexity starts quadratically,
in agreement with eq.~\eqref{eq:universal-short-time}. If
$\mathcal M'_J(\theta_0)=0$, the result reduces to the constant-$\theta$ branch
$|P_y^{(J)}|=(\mathcal E_J/l)|t|$. 

\begingroup
\section{Fixed-charge Routhian treatment for conserved charges}\label{appendix-fixedcharge-routhian}

When a probe moves both radially in AdS and along an internal direction, one may define an
unreduced generalised momentum that includes every semiclassical velocity. A cyclic internal
coordinate, however, is conjugate to a conserved charge and must not be treated as an
independent spreading direction inside a fixed-charge sector. The appropriate reduced
functional is the Routhian, obtained by Legendre transforming with respect to the cyclic
velocity before defining the proper momentum. This appendix collects details used in
sections \ref{CONSRSR} and \ref{complexitygiant}, following the fixed-charge logic of
\cite{Chatzis:2026ekd}.

Let us start with a generic one-dimensional radial Routhian of the form
\begin{equation}
{\mathfrak R}(r,\dot r;q)= {\mathfrak g}(r;q)
\sqrt{\frac{r^2}{l^2}-\frac{l^2\dot r^{2}}{r^2}}\,,
\label{eq:routh-generic-form}
\end{equation}
where $q$ denotes one or several fixed charges. The variable associated with the radial proper distance is defined from
\begin{equation}
\dot y=\frac{l\dot r}{r},
\qquad
{\mathfrak R}= {\mathfrak g}(r;q)
\sqrt{\frac{r^2}{l^2}-\dot y^{2}} .
\label{eq:fixed-proper-coordinate-generic}
\end{equation}
The fixed-charge proper momentum is therefore
\begin{equation}
P^{(q)}_y=\frac{\partial {\mathfrak R}}{\partial \dot y}
= - \frac{{\mathfrak g}(r;q)\dot y}{\sqrt{\frac{r^2}{l^2}-\dot y^{2}}} .
\label{eq:fixed-proper-momentum-generic}
\end{equation}
This is the quantity proposed to represent, up to the positive normalisation $\kappa$, the complexity rate in a fixed-charge sector. A velocity along the cyclic variable no longer appears in $\dot y$; its effect survives through the fixed parameter $q$ in ${\mathfrak g}(r;q)$ and in the conserved energy. The function ${\mathfrak g}$ may be constant or may depend on the remaining reduced coordinates.

For the point particle of section \ref{CONSRSR},
\begin{equation}
L_p=-m\sqrt{\frac{r^2}{l^2}-\frac{l^2\dot r^2}{r^2}-l^2\dot\psi^{2}},
\qquad
J=\frac{\partial L_p}{\partial \dot\psi} .
\label{eq:routh-point-lag}
\end{equation}
Solving the first relation for $\dot\psi(r,\dot r,J)$ and Legendre transforming gives
\begin{equation}
R_J=J\dot\psi-L_p=\frac{\sqrt{J^2+m^2l^2}}{l}
\sqrt{\frac{r^2}{l^2}-\frac{l^2\dot r^2}{r^2}}
\equiv \widehat m
\sqrt{\frac{r^2}{l^2}-\frac{l^2\dot r^2}{r^2}},
\qquad
\widehat m^2=\frac{J^2+m^2l^2}{l^2} .
\label{eq:routh-point-result}
\end{equation}
The charge $J$ has become part of the 'effective mass' $\hat{m}$. The Routhian-derived  Hamiltonian is
\begin{equation}
H_J=P_r\dot r-R_J
= -\frac{\widehat m r^2/l^2}{\sqrt{\frac{r^2}{l^2}-\frac{l^2\dot r^2}{r^2}}}
\label{eq:routh-point-H}
\end{equation}
up to the same sign convention used in the main text. Hence
\begin{equation}
\dot r=\pm\frac{r^2}{H_J l^4}
\sqrt{H_J^2l^4-\widehat m^2l^2 r^2},
\qquad
r(t)=\frac{|H_J|l^2}{\sqrt{H_J^2t^2+\widehat m^2l^2}} .
\label{eq:routh-point-solution}
\end{equation}
Using $\dot y=l\dot r/r$ in eq.~\eqref{eq:fixed-proper-momentum-generic}, one obtains
\begin{equation}
P^{(J)}_y(t)= -\frac{H_J}{l}\,t ,
\qquad
\dot C_J(t)\propto |P^{(J)}_y(t)|=\frac{|H_J|}{l}|t| .
\label{eq:routh-point-complexity-rate}
\end{equation}
In particular $P^{(J)}_y(0)=0$. This is the desired fixed-charge short-time behaviour: the complexity itself starts as $C_J(t)-C_J(0)\sim t^2$ \cite{Muck:2026top}. The constant term in eq.~\eqref{cdotatsmallt} is therefore not a fixed-$J$ Krylov growth rate; it is the contribution of the non-zero internal velocity $\dot\psi(0)$ in the unreduced configuration space.

We now apply the same logic to the giant graviton of section \ref{complexitygiant}. Its
Born--Infeld plus Wess--Zumino Lagrangian is
\begin{equation}
L_{\rm giant}=\nu\left[\cos^3\theta
\sqrt{\frac{r^2}{l^2}-\frac{l^2\dot r^2}{r^2}-l^2\dot\theta^2
-l^2\sin^2\theta\dot\phi^2}
-\mu\cos^4\theta\dot\phi\right].
\label{eq:routh-giant-lag}
\end{equation}
The unsquared relation $J=\partial L_{\rm giant}/\partial\dot\phi$ fixes the physical
solution \eqref{phidot}. With $Q(\theta)=J+\mu\nu\cos^4\theta$, the partial Legendre
transform is
\begin{equation}
R_J=J\dot\phi-L_{\rm giant}=-\mathcal M_J(\theta)
\sqrt{\frac{r^2}{l^2}-\frac{l^2\dot r^2}{r^2}-l^2\dot\theta^2},
\qquad
\mathcal M_J(\theta)=
\frac{\sqrt{Q^2(\theta)+\nu^2l^2\sin^2\theta\cos^6\theta}}
{l\sin\theta}.
\label{eq:routh-giant-R}
\end{equation}
Using $\widetilde R_J=-R_J$ leaves the reduced equations unchanged and gives a positive
effective mass. The Wess--Zumino coupling is encoded in the shifted charge $Q(\theta)$.
\\
The reduced proper coordinate and momentum are
\begin{align}
\dot y_J^2&=\frac{l^2\dot r^2}{r^2}+l^2\dot\theta^2,
\qquad
\widetilde R_J=\mathcal M_J(\theta)
\sqrt{\frac{r^2}{l^2}-\dot y_J^2},
\label{eq:routh-giant-proper-coordinate}\\
P_y^{(J)}&=-\frac{\mathcal M_J(\theta)\dot y_J}
{\sqrt{r^2/l^2-\dot y_J^2}}
=-\frac{\mathcal E_Jl^2}{r^2}\dot y_J.
\label{eq:routh-giant-Py}
\end{align}
No $\phi$ velocity is reintroduced. For the rest data
\eqref{eq:giant-reduced-rest-data}, appendix \ref{appendixc} proves that
$P_y^{(J)}(0)=0$ and $\dot{\mathcal C}_J\sim t$. A trajectory with a prescribed
$\dot\theta(0)\neq0$ is a different bulk preparation and lies outside the standard
Krylov identification about that reference time; it is not used to evade the universal
condition $\dot{\mathcal C}_K(0)=0$.
\endgroup

\bibliographystyle{JHEP}
\bibliography{main}

@article{Huh:2023xtr,
    author = "Huh, Kyoung-Bum and Jeong, Hyun-Sik and Pedraza, Juan F.",
    title = "{Spread complexity in saddle-dominated scrambling}",
    eprint = "2312.12593",
    archivePrefix = "arXiv",
    primaryClass = "hep-th",
    reportNumber = "IFT-UAM/CSIC-23-178",
    doi = "10.1007/JHEP05(2024)137",
    journal = "JHEP",
    volume = "05",
    pages = "137",
    year = "2024"
}

@article{Fan:2022xaa,
    author = "Fan, Zhong-Ying",
    title = "{Universal relation for operator complexity}",
    eprint = "2202.07220",
    archivePrefix = "arXiv",
    primaryClass = "quant-ph",
    doi = "10.1103/PhysRevA.105.062210",
    journal = "Phys. Rev. A",
    volume = "105",
    number = "6",
    pages = "062210",
    year = "2022"
}

@article{Balasubramanian:2025xkj,
    author = "Balasubramanian, Vijay and Caputa, Pawel and Sim{\'o}n, Joan",
    title = "{Variations on a theme of Krylov}",
    eprint = "2511.03775",
    archivePrefix = "arXiv",
    primaryClass = "hep-th",
    reportNumber = "YITP-25-171",
    doi = "10.1007/JHEP04(2026)172",
    journal = "JHEP",
    volume = "04",
    pages = "172",
    year = "2026"
}

@article{Muck:2026top,
    author = {M{\"u}ck, Wolfgang},
    title = "{Krylov complexity has it all}",
    eprint = "2605.28681",
    archivePrefix = "arXiv",
    primaryClass = "hep-th",
    month = "5",
    year = "2026"
}

@article{Chatzis:2026ekd,
    author = "Chatzis, Dimitrios and Hammond, Madison and Nunez, Carlos and Ramallo, Alfonso V. and Santamaria, Ricardo T.",
    title = "{Holographic Spread Complexity from Branes and Strings}",
    eprint = "2607.00074",
    archivePrefix = "arXiv",
    primaryClass = "hep-th",
    month = "6",
    year = "2026"
}

@article{Anabalon:2021tua,
    author = "Anabalon, Andres and Ross, Simon F.",
    title = "{Supersymmetric solitons and a degeneracy of solutions in AdS/CFT}",
    eprint = "2104.14572",
    archivePrefix = "arXiv",
    primaryClass = "hep-th",
    doi = "10.1007/JHEP07(2021)015",
    journal = "JHEP",
    volume = "07",
    pages = "015",
    year = "2021"
}

@article{Nunez:2023nnl,
    author = "Nunez, Carlos and Oyarzo, Marcelo and Stuardo, Ricardo",
    title = "{Confinement in (1 + 1) dimensions: a holographic perspective from I-branes}",
    eprint = "2307.04783",
    archivePrefix = "arXiv",
    primaryClass = "hep-th",
    doi = "10.1007/JHEP09(2023)201",
    journal = "JHEP",
    volume = "09",
    pages = "201",
    year = "2023"
}

@article{Anabalon:2023lnk,
    author = "Anabal\'on, Andr\'es and Nastase, Horatiu",
    title = "{Universal IR holography, scalar fluctuations, and glueball spectra}",
    eprint = "2310.07823",
    archivePrefix = "arXiv",
    primaryClass = "hep-th",
    doi = "10.1103/PhysRevD.109.066011",
    journal = "Phys. Rev. D",
    volume = "109",
    number = "6",
    pages = "066011",
    year = "2024"
}

@article{Nunez:2023xgl,
    author = "Nunez, Carlos and Oyarzo, Marcelo and Stuardo, Ricardo",
    title = "{Confinement and D5-branes}",
    eprint = "2311.17998",
    archivePrefix = "arXiv",
    primaryClass = "hep-th",
    doi = "10.1007/JHEP03(2024)080",
    journal = "JHEP",
    volume = "03",
    pages = "080",
    year = "2024"
}

@article{Anabalon:2024che,
    author = "Anabal\'on, Andr\'es and Nastase, Horatiu and Oyarzo, Marcelo",
    title = "{Supersymmetric AdS Solitons and the interconnection of different vacua of ${\cal N}=4$ Super Yang-Mills}",
    eprint = "2402.18482",
    archivePrefix = "arXiv",
    primaryClass = "hep-th",
    month = "2",
    year = "2024"
}

@article{Chatzis:2024kdu,
    author = "Chatzis, Dimitrios and Fatemiabhari, Ali and Nunez, Carlos and Weck, Peter",
    title = "{SCFT deformations via uplifted solitons}",
    eprint = "2406.01685",
    archivePrefix = "arXiv",
    primaryClass = "hep-th",
    month = "6",
    year = "2024"
}

@article{Chatzis:2024top,
    author = "Chatzis, Dimitrios and Fatemiabhari, Ali and Nunez, Carlos and Weck, Peter",
    title = "{Conformal to confining SQFTs from holography}",
    eprint = "2405.05563",
    archivePrefix = "arXiv",
    primaryClass = "hep-th",
    month = "5",
    year = "2024"
}

@article{Lozano:2020txg,
    author = "Lozano, Yolanda and Nunez, Carlos and Ramirez, Anayeli and Speziali, Stefano",
    title = "{New AdS$_{2}$ backgrounds and $ \mathcal{N} $ = 4 conformal quantum mechanics}",
    eprint = "2011.00005",
    archivePrefix = "arXiv",
    primaryClass = "hep-th",
    doi = "10.1007/JHEP03(2021)277",
    journal = "JHEP",
    volume = "03",
    pages = "277",
    year = "2021"
}

@article{Lozano:2019zvg,
    author = "Lozano, Yolanda and Macpherson, Niall T. and Nunez, Carlos and Ramirez, Anayeli",
    title = "{Two dimensional ${\cal N}=(0,4)$ quivers dual to AdS$_3$ solutions in massive IIA}",
    eprint = "1909.10510",
    archivePrefix = "arXiv",
    primaryClass = "hep-th",
    doi = "10.1007/JHEP01(2020)140",
    journal = "JHEP",
    volume = "01",
    pages = "140",
    year = "2020"
}

@article{Lozano:2020bxo,
    author = "Lozano, Yolanda and Nunez, Carlos and Ramirez, Anayeli and Speziali, Stefano",
    title = "{$M$-strings and AdS$_3$ solutions to M-theory with small $\mathcal{N}=(0,4)$ supersymmetry}",
    eprint = "2005.06561",
    archivePrefix = "arXiv",
    primaryClass = "hep-th",
    doi = "10.1007/JHEP08(2020)118",
    journal = "JHEP",
    volume = "08",
    pages = "118",
    year = "2020"
}

@article{Caputa:2024sux,
    author = "Caputa, Pawel and Chen, Bowen and McDonald, Ross W. and Sim{\'o}n, Joan and Strittmatter, Benjamin",
    title = "{Spread Complexity Rate as Proper Momentum}",
    eprint = "2410.23334",
    archivePrefix = "arXiv",
    primaryClass = "hep-th",
    reportNumber = "YITP-24-137",
    month = "10",
    year = "2024"
}

@article{Baiguera:2025dkc,
    author = "Baiguera, Stefano and Balasubramanian, Vijay and Caputa, Pawel and Chapman, Shira and Haferkamp, Jonas and Heller, Michal P. and Halpern, Nicole Yunger",
    title = "{Quantum complexity in gravity, quantum field theory, and quantum information science}",
    eprint = "2503.10753",
    archivePrefix = "arXiv",
    primaryClass = "hep-th",
    reportNumber = "YITP-25-39",
    month = "3",
    year = "2025"
}

@article{Balasubramanian:2022tpr,
    author = "Balasubramanian, Vijay and Caputa, Pawel and Magan, Javier M. and Wu, Qingyue",
    title = "{Quantum chaos and the complexity of spread of states}",
    eprint = "2202.06957",
    archivePrefix = "arXiv",
    primaryClass = "hep-th",
    doi = "10.1103/PhysRevD.106.046007",
    journal = "Phys. Rev. D",
    volume = "106",
    number = "4",
    pages = "046007",
    year = "2022"
}

@article{Rabinovici:2025otw,
    author = "Rabinovici, Eliezer and S{\'a}nchez-Garrido, Adri{\'a}n and Shir, Ruth and Sonner, Julian",
    title = "{Krylov Complexity}",
    eprint = "2507.06286",
    archivePrefix = "arXiv",
    primaryClass = "hep-th",
    reportNumber = "CERN-TH-2025-128",
    month = "7",
    year = "2025"
}

@article{Jeong:2026iac,
    author = "Jeong, Hyun-Sik",
    title = "{Krylov Subspace Dynamics as Near-Horizon AdS$_2$ Holography}",
    eprint = "2602.11627",
    archivePrefix = "arXiv",
    primaryClass = "hep-th",
    reportNumber = "APCTP Pre2026 - 003",
    month = "2",
    year = "2026"
}

@article{Zoakos:2026obl,
    author = "Zoakos, Dimitrios",
    title = "{Holographic Krylov complexity in the Coulomb branch of ${\cal N}=4$ SYM}",
    eprint = "2603.15435",
    archivePrefix = "arXiv",
    primaryClass = "hep-th",
    month = "3",
    year = "2026"
}

@article{Li:2025fqz,
    author = "Li, Zhehan and Tian, Jia",
    title = "{The Holography of Spread Complexity: A Story of Observers}",
    eprint = "2506.13481",
    archivePrefix = "arXiv",
    primaryClass = "hep-th",
    month = "6",
    year = "2025"
}

@article{Anabalon:2026yxk,
    author = "Anabal{\'o}n, Andr{\'e}s and Nastase, Horatiu and Nunez, Carlos and Oyarzo, Marcelo and Stuardo, Ricardo",
    title = "{Moduli space of ${\cal N}=4$ Super Yang-Mills from AdS/CFT}",
    eprint = "2603.18141",
    archivePrefix = "arXiv",
    primaryClass = "hep-th",
    month = "3",
    year = "2026"
}

@article{Baggioli:2024wbz,
    author = "Baggioli, Matteo and Huh, Kyoung-Bum and Jeong, Hyun-Sik and Kim, Keun-Young and Pedraza, Juan F.",
    title = "{Krylov complexity as an order parameter for quantum chaotic-integrable transitions}",
    eprint = "2407.17054",
    archivePrefix = "arXiv",
    primaryClass = "hep-th",
    reportNumber = "IFT-UAM/CSIC-24-107",
    doi = "10.1103/PhysRevResearch.7.023028",
    journal = "Phys. Rev. Res.",
    volume = "7",
    number = "2",
    pages = "023028",
    year = "2025"
}

@article{Caputa:2021sib,
    author = "Caputa, Pawel and Magan, Javier M. and Patramanis, Dimitrios",
    title = "{Geometry of Krylov complexity}",
    eprint = "2109.03824",
    archivePrefix = "arXiv",
    primaryClass = "hep-th",
    doi = "10.1103/PhysRevResearch.4.013041",
    journal = "Phys. Rev. Res.",
    volume = "4",
    number = "1",
    pages = "013041",
    year = "2022"
}

@article{Fan:2024iop,
    author = "Fan, Zhong-Ying",
    title = "{Momentum-Krylov complexity correspondence}",
    eprint = "2411.04492",
    archivePrefix = "arXiv",
    primaryClass = "hep-th",
    month = "11",
    year = "2024"
}

@article{He:2024pox,
    author = "He, Peng-Zhang",
    title = "{Revisit the relationship between spread complexity rate and radial momentum}",
    eprint = "2411.19172",
    archivePrefix = "arXiv",
    primaryClass = "hep-th",
    month = "11",
    year = "2024"
}

@article{Rabinovici:2023yex,
    author = "Rabinovici, E. and S{\'a}nchez-Garrido, A. and Shir, R. and Sonner, J.",
    title = "{A bulk manifestation of Krylov complexity}",
    eprint = "2305.04355",
    archivePrefix = "arXiv",
    primaryClass = "hep-th",
    doi = "10.1007/JHEP08(2023)213",
    journal = "JHEP",
    volume = "08",
    pages = "213",
    year = "2023"
}

@article{Fatemiabhari:2025cyy,
    author = "Fatemiabhari, Ali and Nastase, Horatiu and Roychowdhury, Dibakar",
    title = "{Holographic Krylov complexity in ${\cal N}=4$ SYM}",
    eprint = "2511.19286",
    archivePrefix = "arXiv",
    primaryClass = "hep-th",
    month = "11",
    year = "2025"
}

@article{Chatzis:2025hek,
    author = "Chatzis, Dimitrios and Hammond, Madison and Itsios, Georgios and Nunez, Carlos and Zoakos, Dimitrios",
    title = "{Supersymmetric AdS Solitons, Coulomb Branch Flows and Twisted Compactifications}",
    eprint = "2511.18128",
    archivePrefix = "arXiv",
    primaryClass = "hep-th",
    month = "11",
    year = "2025"
}

@article{Lozano:2019emq,
    author = "Lozano, Yolanda and Macpherson, Niall T. and Nunez, Carlos and Ramirez, Anayeli",
    title = "{AdS$_3$ solutions in Massive IIA with small $\mathcal{N}=(4,0)$ supersymmetry}",
    eprint = "1908.09851",
    archivePrefix = "arXiv",
    primaryClass = "hep-th",
    doi = "10.1007/JHEP01(2020)129",
    journal = "JHEP",
    volume = "01",
    pages = "129",
    year = "2020"
}

@article{Lozano:2019jza,
    author = "Lozano, Yolanda and Macpherson, Niall T. and Nunez, Carlos and Ramirez, Anayeli",
    title = "{1/4 BPS solutions and the AdS$_3$/CFT$_2$ correspondence}",
    eprint = "1909.09636",
    archivePrefix = "arXiv",
    primaryClass = "hep-th",
    doi = "10.1103/PhysRevD.101.026014",
    journal = "Phys. Rev. D",
    volume = "101",
    number = "2",
    pages = "026014",
    year = "2020"
}

@article{Fatemiabhari:2025usn,
    author = "Fatemiabhari, Ali and Nastase, Horatiu and Nunez, Carlos and Roychowdhury, Dibakar",
    title = "{Holographic Krylov complexity in confining gauge theories}",
    eprint = "2511.22717",
    archivePrefix = "arXiv",
    primaryClass = "hep-th",
    month = "11",
    year = "2025"
}

@article{Das:2024tnw,
    author = "Das, Rathindra Nath and Demulder, Saskia and Erdmenger, Johanna and Northe, Christian",
    title = "{Spread complexity for the planar limit of holography}",
    eprint = "2412.09673",
    archivePrefix = "arXiv",
    primaryClass = "hep-th",
    doi = "10.1007/JHEP06(2025)166",
    journal = "JHEP",
    volume = "06",
    pages = "166",
    year = "2025"
}

@article{Fatemiabhari:2025poq,
    author = "Fatemiabhari, Ali and Nastase, Horatiu and Nunez, Carlos and Roychowdhury, Dibakar",
    title = "{Holographic Krylov Complexity for Conformal Quiver Gauge Theories}",
    eprint = "2512.14812",
    archivePrefix = "arXiv",
    primaryClass = "hep-th",
    month = "12",
    year = "2025"
}

@article{Roychowdhury:2026eds,
    author = "Roychowdhury, Dibakar",
    title = "{Holographic Krylov complexity for Yang-Baxter deformed supergravity backgrounds}",
    eprint = "2601.06555",
    archivePrefix = "arXiv",
    primaryClass = "hep-th",
    month = "1",
    year = "2026"
}

@article{Fatemiabhari:2026goj,
    author = "Fatemiabhari, Ali and Nunez, Carlos",
    title = "{Krylov Complexity, Confinement and Universality}",
    eprint = "2602.17757",
    archivePrefix = "arXiv",
    primaryClass = "hep-th",
    month = "2",
    year = "2026"
}

@article{Witten:1998xy,
    author = "Witten, Edward",
    title = "{Baryons and branes in anti-de Sitter space}",
    eprint = "hep-th/9805112",
    archivePrefix = "arXiv",
    reportNumber = "IASSNS-HEP-98-42",
    doi = "10.1088/1126-6708/1998/07/006",
    journal = "JHEP",
    volume = "07",
    pages = "006",
    year = "1998"
}

@article{Berenstein:2002jq,
    author = "Berenstein, David Eliecer and Maldacena, Juan Martin and Nastase, Horatiu Stefan",
    title = "{Strings in flat space and pp waves from N=4 superYang-Mills}",
    eprint = "hep-th/0202021",
    archivePrefix = "arXiv",
    doi = "10.1088/1126-6708/2002/04/013",
    journal = "JHEP",
    volume = "04",
    pages = "013",
    year = "2002"
}

@article{Caputa:2025ozd,
    author = "Caputa, Pawel and Di Giulio, Giuseppe and Loc, Tran Quang",
    title = "{Symmetry-resolved spread complexity}",
    eprint = "2509.12992",
    archivePrefix = "arXiv",
    primaryClass = "hep-th",
    reportNumber = "YITP-25-146",
    doi = "10.1007/JHEP02(2026)189",
    journal = "JHEP",
    volume = "02",
    pages = "189",
    year = "2026"
}

@article{Caputa:2025mii,
    author = "Caputa, Pawel and Di Giulio, Giuseppe and Loc, Tran Quang",
    title = "{Growth of block-diagonal operators and symmetry-resolved Krylov complexity}",
    eprint = "2507.02033",
    archivePrefix = "arXiv",
    primaryClass = "hep-th",
    reportNumber = "YITP-25-101",
    doi = "10.1103/9v9v-54zv",
    journal = "Phys. Rev. Res.",
    volume = "7",
    number = "4",
    pages = "043055",
    year = "2025"
}

@article{Craps:2024suj,
    author = "Craps, Ben and Evnin, Oleg and Pascuzzi, Gabriele",
    title = "{Multiseed Krylov Complexity}",
    eprint = "2409.15666",
    archivePrefix = "arXiv",
    primaryClass = "quant-ph",
    doi = "10.1103/PhysRevLett.134.050402",
    journal = "Phys. Rev. Lett.",
    volume = "134",
    number = "5",
    pages = "050402",
    year = "2025"
}

@article{PG:2025ixk,
    author = "PG, Sreeram and Kannan, J. Bharathi and Modak, Ranjan and Aravinda, S.",
    title = "{Dependence of Krylov complexity saturation on the initial operator and state}",
    eprint = "2503.03400",
    archivePrefix = "arXiv",
    primaryClass = "quant-ph",
    doi = "10.1103/hsvm-w849",
    journal = "Phys. Rev. E",
    volume = "112",
    number = "3",
    pages = "L032203",
    year = "2025"
}

@article{Callan:1998iq,
    author = "Callan, Jr., Curtis G. and Guijosa, Alberto and Savvidy, Konstantin G.",
    title = "{Baryons and string creation from the five-brane world volume action}",
    eprint = "hep-th/9810092",
    archivePrefix = "arXiv",
    reportNumber = "PUPT-1814",
    doi = "10.1016/S0550-3213(99)00057-7",
    journal = "Nucl. Phys. B",
    volume = "547",
    pages = "127--142",
    year = "1999"
}

@article{Gomis:1999xs,
    author = "Gomis, Joaquim and Ramallo, Alfonso V. and Simon, Joan and Townsend, Paul K.",
    title = "{Supersymmetric baryonic branes}",
    eprint = "hep-th/9907022",
    archivePrefix = "arXiv",
    reportNumber = "UB-ECM-PF-99-09, US-FT-14-99, DAMTP-1999-53",
    doi = "10.1088/1126-6708/1999/11/019",
    journal = "JHEP",
    volume = "11",
    pages = "019",
    year = "1999"
}

@article{Janssen:2006sc,
    author = "Janssen, Bert and Lozano, Yolanda and Rodriguez-Gomez, Diego",
    title = "{The Baryon vertex with magnetic flux}",
    eprint = "hep-th/0606264",
    archivePrefix = "arXiv",
    reportNumber = "UG-FT-203-06, CAFPE-73-06, FFUOV-06-09, FTUAM-06-07",
    doi = "10.1088/1126-6708/2006/11/082",
    journal = "JHEP",
    volume = "11",
    pages = "082",
    year = "2006"
}

@article{Camino:2001at,
    author = "Camino, J. M. and Paredes, Angel and Ramallo, A. V.",
    title = "{Stable wrapped branes}",
    eprint = "hep-th/0104082",
    archivePrefix = "arXiv",
    reportNumber = "US-FT-3-01",
    doi = "10.1088/1126-6708/2001/05/011",
    journal = "JHEP",
    volume = "05",
    pages = "011",
    year = "2001"
}

@article{Caldarelli:2004yk,
    author = "Caldarelli, Marco M. and Silva, Pedro J.",
    title = "{Multi-giant graviton systems, SUSY breaking and CFT}",
    eprint = "hep-th/0401213",
    archivePrefix = "arXiv",
    reportNumber = "IFUM-785-FT",
    doi = "10.1088/1126-6708/2004/02/052",
    journal = "JHEP",
    volume = "02",
    pages = "052",
    year = "2004"
}

@article{McGreevy:2000cw,
    author = "McGreevy, John and Susskind, Leonard and Toumbas, Nicolaos",
    title = "{Invasion of the giant gravitons from Anti-de Sitter space}",
    eprint = "hep-th/0003075",
    archivePrefix = "arXiv",
    reportNumber = "SU-ITP-00-09",
    doi = "10.1088/1126-6708/2000/06/008",
    journal = "JHEP",
    volume = "06",
    pages = "008",
    year = "2000"
}

@article{Fatemiabhari:2026rob,
    author = "Fatemiabhari, Ali and Nunez, Carlos and Santamaria, Ricardo T.",
    title = "{Complexity and Operator Growth in Holographic 6d SCFTs}",
    eprint = "2603.10106",
    archivePrefix = "arXiv",
    primaryClass = "hep-th",
    month = "3",
    year = "2026"
}

@article{Parker:2018yvk,
    author = "Parker, Daniel E. and Cao, Xiangyu and Avdoshkin, Alexander and Scaffidi, Thomas and Altman, Ehud",
    title = "{A Universal Operator Growth Hypothesis}",
    eprint = "1812.08657",
    archivePrefix = "arXiv",
    primaryClass = "cond-mat.stat-mech",
    doi = "10.1103/PhysRevX.9.041017",
    journal = "Phys. Rev. X",
    volume = "9",
    number = "4",
    pages = "041017",
    year = "2019"
}

@article{Barbon:2019wsy,
    author = "Barb{\'o}n, J. L. F. and Rabinovici, E. and Shir, R. and Sinha, R.",
    title = "{On The Evolution Of Operator Complexity Beyond Scrambling}",
    eprint = "1907.05393",
    archivePrefix = "arXiv",
    primaryClass = "hep-th",
    reportNumber = "IFT-UAM/CSIC-19-98",
    doi = "10.1007/JHEP10(2019)264",
    journal = "JHEP",
    volume = "10",
    pages = "264",
    year = "2019"
}

@article{Avdoshkin:2019trj,
    author = "Avdoshkin, Alexander and Dymarsky, Anatoly",
    title = "{Euclidean operator growth and quantum chaos}",
    eprint = "1911.09672",
    archivePrefix = "arXiv",
    primaryClass = "cond-mat.stat-mech",
    doi = "10.1103/PhysRevResearch.2.043234",
    journal = "Phys. Rev. Res.",
    volume = "2",
    number = "4",
    pages = "043234",
    year = "2020"
}

@article{Dymarsky:2021bjq,
    author = "Dymarsky, Anatoly and Smolkin, Michael",
    title = "{Krylov complexity in conformal field theory}",
    eprint = "2104.09514",
    archivePrefix = "arXiv",
    primaryClass = "hep-th",
    doi = "10.1103/PhysRevD.104.L081702",
    journal = "Phys. Rev. D",
    volume = "104",
    number = "8",
    pages = "L081702",
    year = "2021"
}

@article{Nandy:2024evd,
    author = "Nandy, Pratik and Matsoukas-Roubeas, Apollonas S. and Mart{\'\i}nez-Azcona, Pablo and Dymarsky, Anatoly and del Campo, Adolfo",
    title = "{Quantum dynamics in Krylov space: Methods and applications}",
    eprint = "2405.09628",
    archivePrefix = "arXiv",
    primaryClass = "quant-ph",
    reportNumber = "RIKEN-iTHEMS-Report-24",
    doi = "10.1016/j.physrep.2025.05.001",
    journal = "Phys. Rept.",
    volume = "1125-1128",
    pages = "1--82",
    year = "2025"
}

@article{Rabinovici:2020ryf,
    author = "Rabinovici, E. and S{\'a}nchez-Garrido, A. and Shir, R. and Sonner, J.",
    title = "{Operator complexity: a journey to the edge of Krylov space}",
    eprint = "2009.01862",
    archivePrefix = "arXiv",
    primaryClass = "hep-th",
    doi = "10.1007/JHEP06(2021)062",
    journal = "JHEP",
    volume = "06",
    pages = "062",
    year = "2021"
}

@article{Rabinovici:2022beu,
    author = "Rabinovici, E. and S{\'a}nchez-Garrido, A. and Shir, R. and Sonner, J.",
    title = "{Krylov complexity from integrability to chaos}",
    eprint = "2207.07701",
    archivePrefix = "arXiv",
    primaryClass = "hep-th",
    doi = "10.1007/JHEP07(2022)151",
    journal = "JHEP",
    volume = "07",
    pages = "151",
    year = "2022"
}

@article{Susskind:2014rva,
    author = "Susskind, Leonard",
    title = "{Computational Complexity and Black Hole Horizons}",
    eprint = "1403.5695",
    archivePrefix = "arXiv",
    primaryClass = "hep-th",
    doi = "10.1002/prop.201500092",
    journal = "Fortsch. Phys.",
    volume = "64",
    pages = "24--43",
    year = "2016",
    note = "[Addendum: Fortsch.Phys. 64, 44--48 (2016)]"
}

@article{Brown:2015bva,
    author = "Brown, Adam R. and Roberts, Daniel A. and Susskind, Leonard and Swingle, Brian and Zhao, Ying",
    title = "{Holographic Complexity Equals Bulk Action?}",
    eprint = "1509.07876",
    archivePrefix = "arXiv",
    primaryClass = "hep-th",
    doi = "10.1103/PhysRevLett.116.191301",
    journal = "Phys. Rev. Lett.",
    volume = "116",
    number = "19",
    pages = "191301",
    year = "2016"
}

@article{Grisaru:2000zn,
    author = "Grisaru, Marcus T. and Myers, Robert C. and Tafjord, Oyvind",
    title = "{SUSY and goliath}",
    eprint = "hep-th/0008015",
    archivePrefix = "arXiv",
    reportNumber = "MCGILL-00-21, BRX-TH-472",
    doi = "10.1088/1126-6708/2000/08/040",
    journal = "JHEP",
    volume = "08",
    pages = "040",
    year = "2000"
}
\end{document}